\documentclass[iop]{emulateapj}
\usepackage{graphicx}
\usepackage{subfigure}
\usepackage{epsfig}
\usepackage{aas_macros}

\newcommand{\swift}{{\it Swift }}
\newcommand{\g}{\gamma}

\def\eBf2{\epsilon_{Bf,-2}}
\def\eef1{\epsilon_{ef,-1}}
\def\RB{{\cal R}_B}
\def\RBbar{\bar{\cal R}_B}

\def\g{\gamma}
\def\G{\Gamma}

\def\nuo{\nu_{\rm opt}}

\def\tx{t_\times}
\def\Ff{F_{\nu}^{f}}
\def\Fr{F_{\nu}^{r}}
\def\tfp{t^{f}_{p}}

\begin{document}

\title{A morphological analysis of gamma-ray burst early optical afterglows}

\author{He Gao$^{1,2,3}$, Xiang-Gao Wang$^{4}$, Peter M\'esz\'aros$^{1,2,3}$, Bing Zhang$^{4}$}
\affil{$^1$Department of Astronomy and Astrophysics, Pennsylvania State University, 525 Davey Laboratory, 
University Park, PA 16802: hug18@psu.edu\\
$^2$Department of Physics, Pennsylvania State University, 525 Davey Laboratory, University Park, PA 
16802\\
$^3$Center for Particle and Gravitational Astrophysics, Institute for Gravitation and the Cosmos, 
Pennsylvania State University, 525 Davey Laboratory, University Park, PA 16802\\
$^4$Department of Physics and Astronomy, University of Nevada Las Vegas, NV 89154, USA}

\begin{abstract}
Within the framework of the external shock model of gamma-ray bursts (GRBs) afterglows, we perform a morphological analysis of the early optical lightcurves to directly constrain model parameters.  We define four morphological types, i.e. the reverse shock dominated cases with/without the emergence of the forward shock peak (Type I/ Type II), and the forward shock dominated cases without/with $\nu_m$ crossing the band (Type III/IV). We systematically investigate all the Swift GRBs that have optical detection earlier than 500 s and find 3/63 Type I bursts ($4.8\%$), 12/63 Type II bursts ($19.0\%$), 30/63 Type III bursts ($47.6\%$), 8/63 Type IV bursts ($12.7\%$) and 10/63 Type III/IV bursts ($15.9\%$).  We perform Monte Carlo simulations to constrain model parameters in order to reproduce the observations. We find that the favored value of the magnetic equipartition parameter in the forward shock ($\epsilon_B^f$) ranges from $10^{-6}$ to $10^{-2}$, and the reverse-to-forward ratio of $\epsilon_B$ ($\RB$) is about 100. The preferred electron equipartition parameter $\epsilon_e^{r,f}$ value is 0.01, which is smaller than the commonly assumed value, e.g., 0.1. This could mitigate the so- called ``efficiency problem" for the internal shock model, if $\epsilon_e$ during the prompt emission phase (in the internal shocks) is large (say, $\sim 0.1$). The preferred $\RB$ value is in agreement with the results in previous works that indicates a moderately magnetized baryonic jet for GRBs. 
\end{abstract}

\section{INTRODUCTION}

The first gamma-ray burst (GRB) afterglow emission was detected in 1997, e.g. X-ray and optical afterglow 
from GRB 970228 \citep{costa97,vanpara97}. Over 18 years a variety of  space- and ground-based 
facilities have detected  hundreds of afterglows, with a wide coverage in both the 
spectral and temporal domains \cite[][for a recent review]{kumarzhang15}.

The standard interpretation for the GRB afterglow emission was proposed before the discovery of 
the first afterglow data \citep{meszarosrees97}. The general picture is as follows 
\cite[][for a review]{gao13review}: regardless of the nature of progenitor and central engine, 
GRBs are believed to originate from a ``fireball" moving at a relativistic speed. The fireball 
will inevitably be decelerated through a pair of shocks (forward and reverse) propagating into 
the ambient medium and the fireball itself. Electrons are accelerated in both shocks and give rise 
to bright non-thermal emission through synchrotron or inverse Compton radiation. Due to the 
deceleration of the fireball, a broad band afterglow emission with power-law rising and decaying behavior is expected for the GRB afterglow.

In the pre-\swift era, the simple external shock model provided successful interpretations for
a large array of afterglow data \citep{wijers97,waxman97a,wijers99,huang99,huang00,
panaitescu01,panaitescu02,yost03}, although moderate revisions were sometimes required, for instance, 
invoking wind-type density medium instead of constant density \citep{dailu98c,meszaros98,
chevalier99,chevalier00}, refining the joint forward shock and reverse shock signal 
\citep{meszarosrees97,meszarosrees99,saripiran99,saripiran99b,kobayashizhang03a,zhang03,wu03,zou05}, 
considering continuous energy injection into the blastwave 
\citep{dailu98b,reesmeszaros98,zhangmeszaros01a}, 
taking into account the jet break effect \citep{rhoads99,sari99,zhangmeszaros02b,rossi02}, etc. 
Entering the \swift era, some new unexpected signatures in GRB afterglows were revealed
\citep{tagliaferri05,burrows05,zhang06,nousek06,obrien06,evans09}, which however are still 
acommodated within the standard framework, provided some additional physical processes are
invoked self-consistently, such as a late central engine activity \citep{zhang06,nousek06}.  

The external shock model is elegant in its simplicity, since it invokes a limited number of 
model parameters (e.g. the total energy of the system, the ambient density and its profile), and has 
well defined predicted spectral and temporal properties. Given this model, the accumulation of afterglow 
data has led to great advances in revealing physical properties in GRB ejecta as well as the circum-burst 
medium. In practice, there are two approaches for applying the external shock model to the observational 
data: one can start with the data, fit the lightcurve and spectrum with some empirical broken 
power-law functions to get both a temporal index $\alpha$ and a spectral index $\beta$, and then 
constrain the related afterglow parameters by applying the so called ``closure relation" 
\citep{zhangmeszaros04,gao13review,wang15}; 
or alternatively, one can start with the external shock model, draw predicted lightcurves and spectrum 
with varying parameters, and then constrain the relevant parameters by fitting the observational data 
with the theoretical prediction. 

Both approaches encounter their own difficulties. The first approach is usually non-optional, since 
some complicated effects such as the equal arrival time effect and the gradual evolution of cooling frequency result in a  smoothing of the spectral and temporal breaks \citep{granotsari02,uhm14b}, leading to imperfection of the ``closure relation". For the latter approach, due to the simple behavior of the afterglow data and the simple power-law property of the synchrotron external shock model, the model parameters obtained by fitting individual bursts usually suffer severe degeneracy, unless the observed SED could fully cover all synchrotron characteristic frequencies (Kumar \& Zhang 2015 for a review), e.g. $\nu_a$ (self-absorption frequency), $\nu_m$ (the characteristic synchrotron frequency of the electrons at the minimum injection energy), and $\nu_c$ (the cooling frequency). For the cases that all the observations are in the same spectral regime or only covers one break frequency, which usually happened when only optical and X-ray data are available, individual fitting cannot make tight constraint on the parameters such as $\epsilon_e$ and $\epsilon_B$, even though the model calculated lightcurve could nicely fit the data (e.g. see the recent results presented in Japelj et al. 2014).

Regardless of these difficulties, both approaches work best with the observational data in the early
stages, which contain much richer information. On the other hand, although multi-wavelength observations 
are routinely carried out for many GRBs, optical data are still generally the most valuable for model 
constraint. The reason is as follows: first, data in the radio band are still limited,  although several radio lightcurves have been interpreted in detail and some interesting studies have been performed statistically \cite[][and reference therein]{chandra12}. Second, the X-ray lightcurves are sometimes dominated by the X-ray flares and X-ray plateaus, which can not be fully interpreted with the simple external shock model, additional physical processes, such as a radiation component that is related to the late central engine activity, are needed \citep{zhang06,nousek06,ghisellini07,kumar08a,
kumar08b}. Last but not least, in the early stage, the reverse shock spectrum is expected to peak in the optical band. Investigation of reverse shock emission is very important for studying the detailed features of GRB ejecta, such as the composition of the jet, since its radiation comes directly from the shocked ejecta materials.

Assuming that afterglow parameters for different GRBs come from the same distributions, statistical properties of a sample of GRBs can be used to constrain the global features of model parameters. For instance,  in the pre-\swift era, \cite{zhang03} proposed to categorize the early optical afterglows into 
different types of combinations of reverse and forward shock emission and they suggested that the afterglow parameter space could be explored based on a morphological analysis. After a decade of successful operation of \swift, a fairly good sample of early afterglow lightcurves is in hand. It is now of great interest to develop and implement the morphological analysis on the current observations to make reasonable constraints on the model parameters. Specifically, the morphological analysis method can be divided into two separate parts: Monte Carlo simulations and observational sample analysis. In the simulation part, we assume some intrinsic distributions for each afterglow parameter, simulate a sample of afterglow lightcurves, and then distribute them into their relevant lightcurve types. In the sample analysis part, we try to find a well defined sample of early optical lightcurves, and calculate the relative number ratios among different lightcurve types. By comparing the results between these two parts, one can make constraints on relevant parameters.

The structure of the paper is as follows. We illustrate the morphological analysis method and the 
sample selection process in section 2, including the definition of different lightcurve categories, 
and the theoretical scheme for determining categories for given values of the afterglow parameters. 
In section 3, we apply the morphological analysis to the GRB sample with Monte Carlo simulations, and 
explore the parameter regimes by comparing the simulation results and observations. We discuss our 
results in section 4, and briefly summarize our conclusions in section 5. Throughout the paper, 
the convention $Q = 10^nQ_n$ is adopted for cgs units.

\begin{figure*}[t]
\subfigure[]{
    \label{fig:subfig:a} %% label for first subfigure
    \includegraphics[width=3.0in]{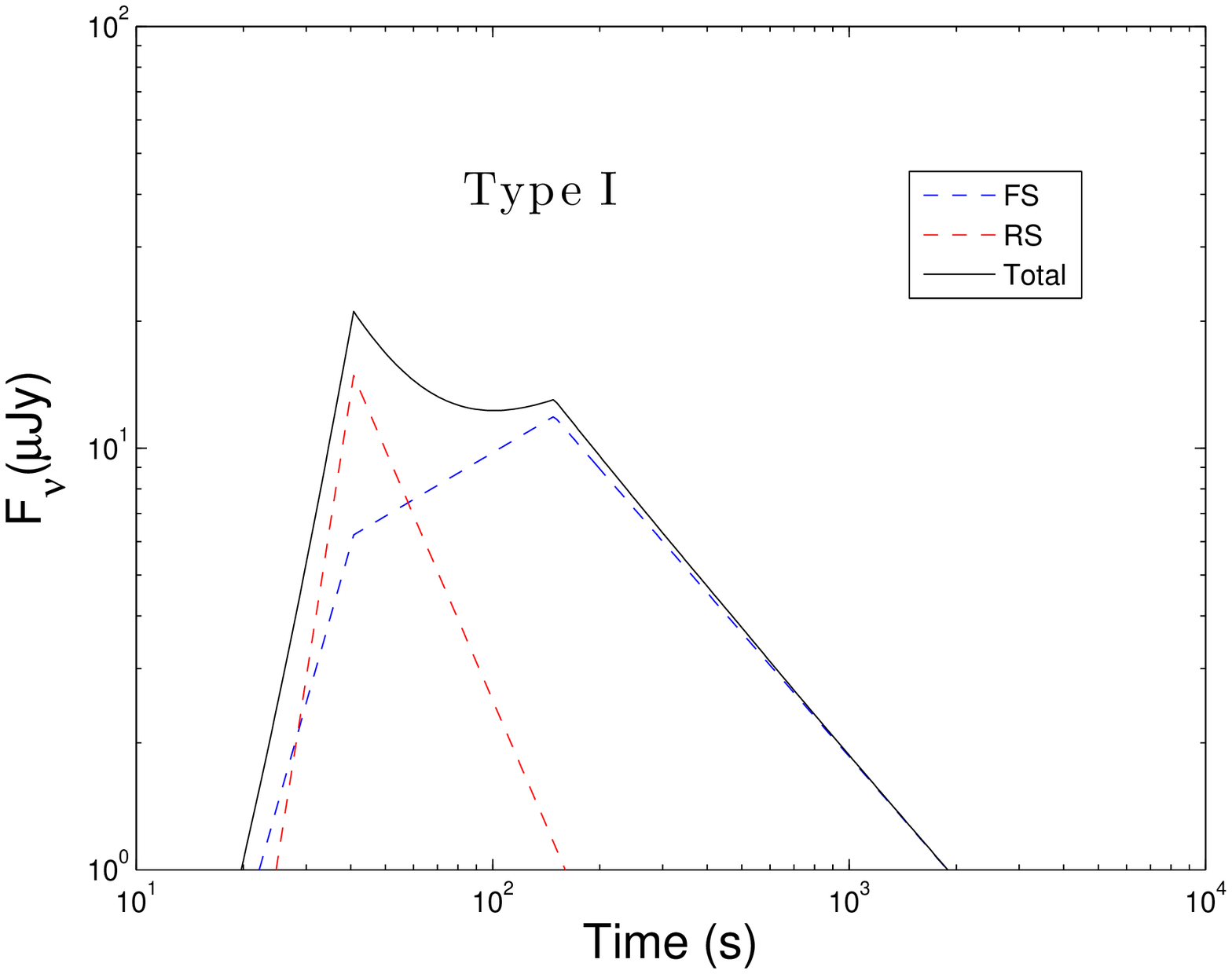}}
    \subfigure[]{
\label{fig:subfig:b} %% label for first subfigure
    \includegraphics[width=3.0in]{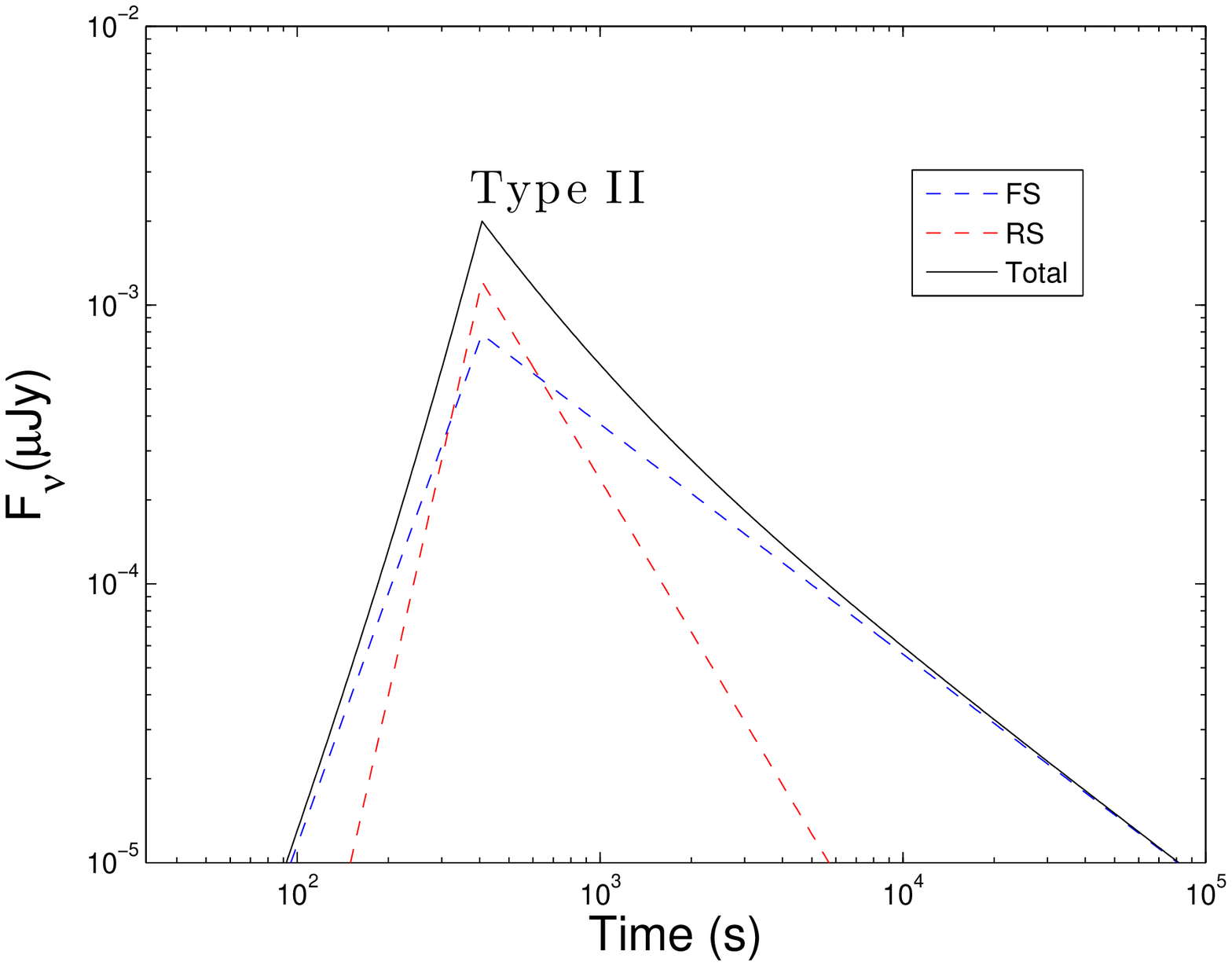}}\\
    \subfigure[]{
    \label{fig:subfig:c} %% label for first subfigure
    \includegraphics[width=3.0in]{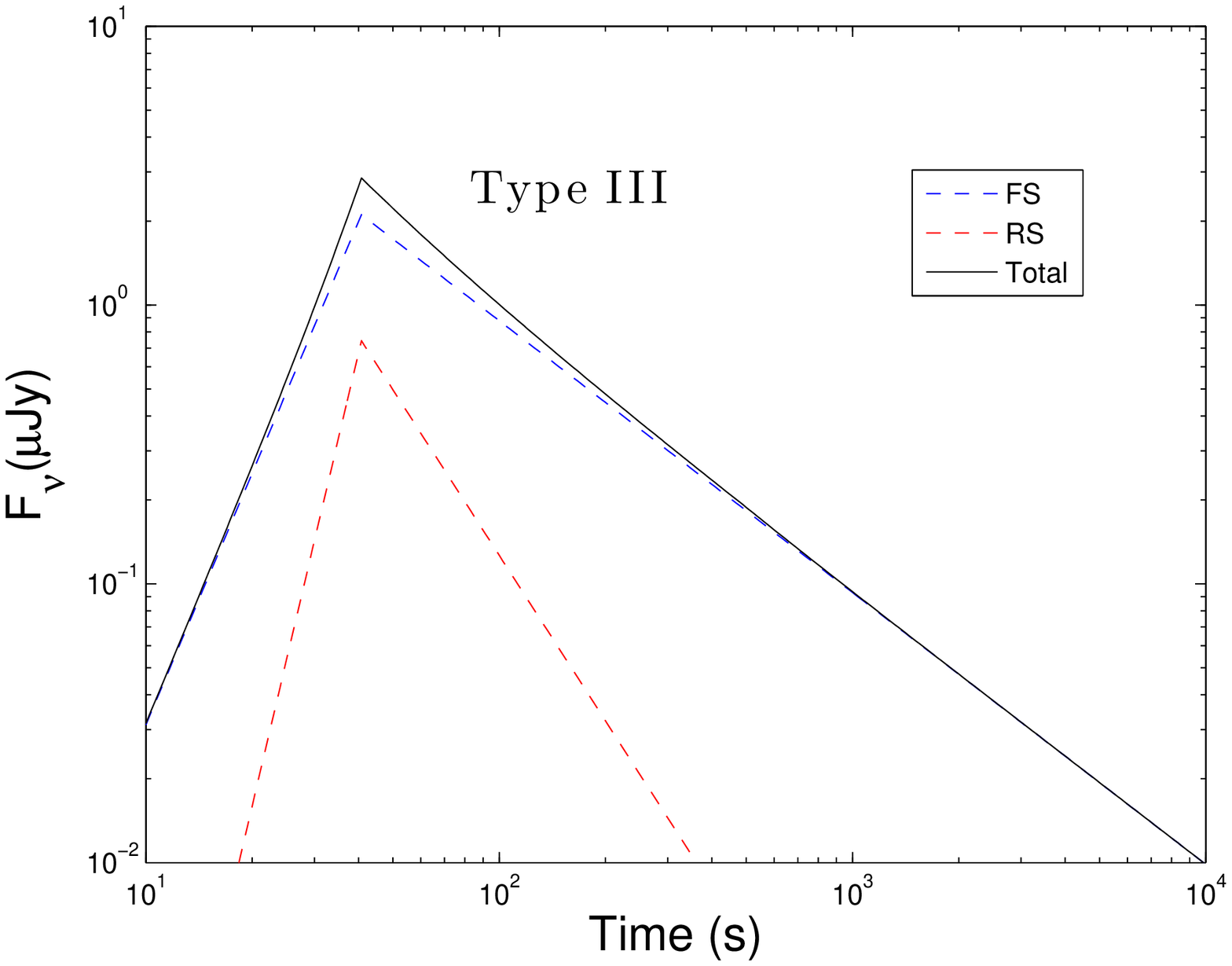}}
    \subfigure[]{
    \centering
    \label{fig:subfig:d} %% label for first subfigure
    \includegraphics[width=3.0in]{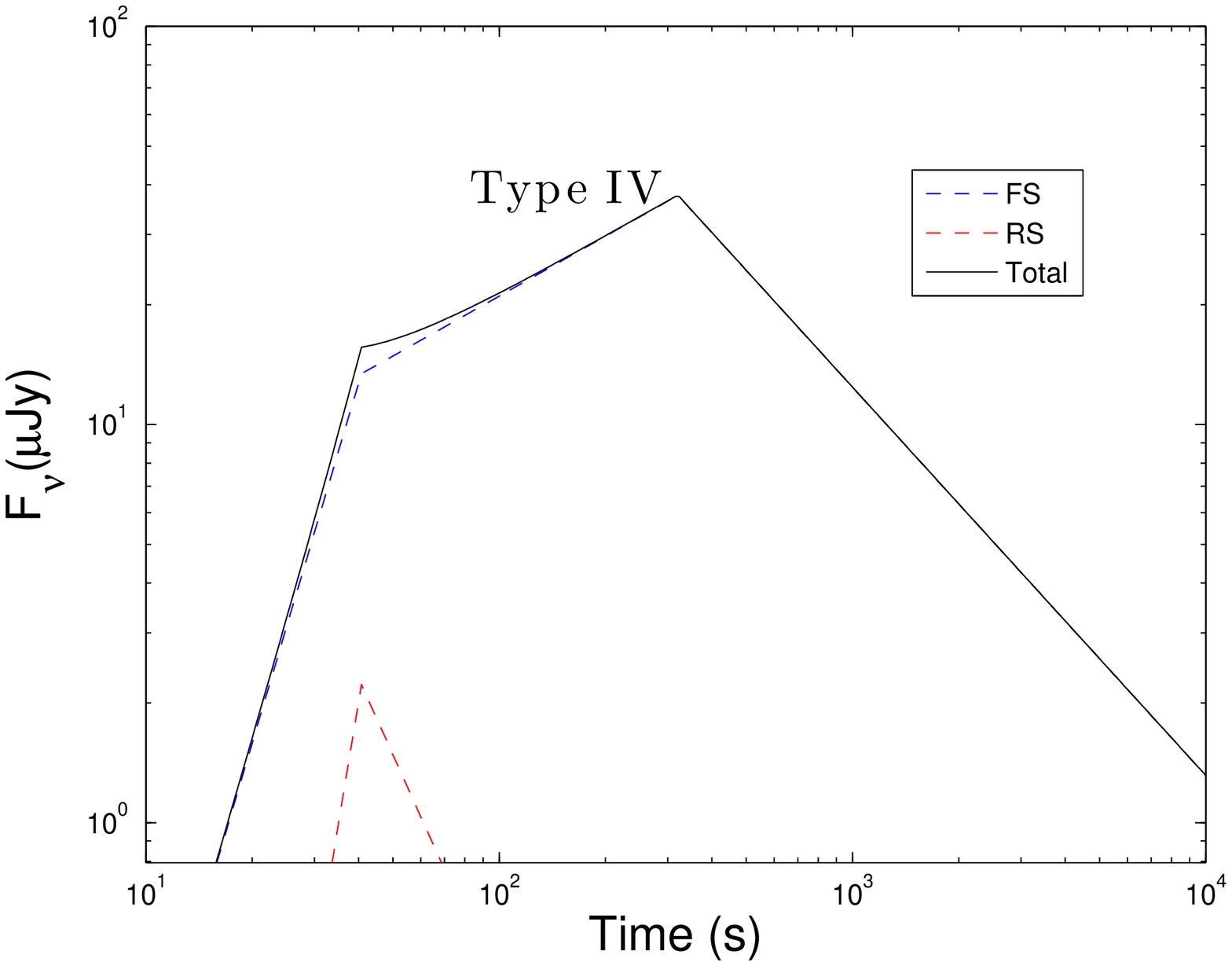}}\\
\caption{{Example light curves for all four types with typical afterglow parameters. Type I: $E=10^{52}~{\rm erg}$, $\Gamma_{0}=100$, $n=10~{\rm cm^{-3}}$, $\epsilon_e^{r,f}=0.1$, $\epsilon_B^{f}=10^{-4}$, $\RB=100$ and $p=2.3$; Type II: $E=10^{52}~{\rm erg}$, $\Gamma_{0}=100$, $n=0.01~{\rm cm^{-3}}$, $\epsilon_e^{r,f}=0.01$, $\epsilon_B^{f}=10^{-5}$, $\RB=100$ and $p=2.1$; Type III: $E=10^{52}~{\rm erg}$, $\Gamma_{0}=100$, $n=10~{\rm cm^{-3}}$, $\epsilon_e^{r,f}=0.01$, $\epsilon_B^{f}=10^{-4}$, $\RB=100$ and $p=2.3$; Type IV: $E=10^{52}~{\rm erg}$, $\Gamma_{0}=100$, $n=10~{\rm cm^{-3}}$, $\epsilon_e^{r,f}=0.1$, $\epsilon_B^{f}=10^{-3}$, $\RB=1$ and $p=2.3$.}}
\label{fig:typeshow}          
\end{figure*}

\section{Early optical afterglow morphology}

\subsection{Lightcurve classification}
The morphology of early optical afterglows essentially reflects the relative relation between the
forward shock and the reverse shock emission. Since the strength of the forward and reverse shocks 
are determined by the same set of GRB parameters, namely the initial Lorentz factor, the kinetic 
energy of the fireball, the circum-burst density and the microphysics parameters, in principle a study of the morphology can yield direct model constraints. 

In previous works, the early optical afterglows {for constant density medium model} were usually classified into three categories \citep{zhang03,jinfan07}:

\begin{itemize}
\item Type I:  re-brightening. At the very early stage, the lightcurve is dominated by the reverse 
shock emission, but later a re-brightening signature emerges due to the forward shock emission 
contribution. Both reverse shock peak and forward shock peak are evident in this type of lightcurve. 

\item Type II: flattening. The forward shock peak is beneath the reverse shock component. 
The forward shock emission only shows its decaying part at the late stage, since the 
reverse shock component fades more rapidly. 

\item Type III: no reverse shock component. In this case, the reverse shock component is either too weak 
compared with the forward shock emission, or is completely suppressed for some reason, such as 
magnetic fields dominating the ejecta \citep{zhangkobayashi05,mimica10}.
\end{itemize}

Note that for forward shock dominated cases (Type III), there are still two distinct shapes of 
lightcurve, depending on whether $\nu_{m}^f(\tx)$ is larger than $\nu_{\rm opt}$ or not, where $\tx$ is the reverse shock crossing time \citep{saripiran95}. If $\nu_{m}^f(\tx)>
\nu_{\rm opt}$, the rising slope of the lightcurve would have a very clear steep ($t^{3/2}$ or $t^{3}$) 
to shallow ($t^{1/2}$) transition, otherwise the rising slope is always steep. Since an insight on
the $\nu_{m}^f(\tx)$ value could lead to strong constraints on relevant afterglow parameters, in this 
work we further categorize the forward shock dominated lightcurves into two categories: 
\begin{itemize}
\item Type III: forward shock dominated lightcurves without $\nu_{m}$ crossing. The observed optical peak is the deceleration peak.
\item Type IV: forward shock dominated lightcurves with $\nu_{m}$ crossing. The observed optical peak is the $\nu_{m}$ crossing peak.
\end{itemize}
Figure \ref{fig:typeshow} shows the example light curves for all four types with typical afterglow parameters.

\subsection{Theoretical scheme for determining categories}
\label{subsec:theo}

Consider a uniform relativistic shell (fireball ejecta) with an isotropic equivalent energy $E$, 
initial Lorentz factor $\G_0$, and observed width $\Delta_0$, expanding into a homogeneous interstellar 
medium (ISM) with particle number density $n$ at a redshift $z$. During the initial interaction, a 
pair of shocks develop: a forward shock propagating into the medium and a reverse shock propagating 
into the shell. After the reverse shock crosses the shell (at $\tx$), the forward shock follows the
Blandford-McKee self-similar solution \citep{blandford76}. {Synchrotron emission is expected 
behind both shocks, since electrons are accelerated at the shock fronts via the 1st-order Fermi 
acceleration mechanism, and magnetic fields are believed to be generated behind the shocks due to plasma instabilities (for forward shock) \citep{medvedev99} or shock compression amplification of the magnetic field carried by the central engine (for reverse shock). }The instantaneous synchrotron spectrum at a given epoch can be 
described with three characteristic frequencies $\nu_a$, $\nu_m$, and $\nu_c$, and the peak synchrotron flux density $F_{\rm{\nu,max}}$ \citep{sari98}. The 
evolution of these four parameters can be calculated, with the help of notations to parameterize the
microscopic processes, i.e. the fractions of shock energy that go to electrons and magnetic fields 
($\epsilon_e$ and $\epsilon_B$), and the electron spectral index $p$. It is then straightforward to 
calculate the lightcurve for a given observed frequency (e.g. optical frequency $\nuo$), 
\begin{equation}
F^{r,f}_{t,\nuo} = f(t,\nuo; z, p, n, \epsilon_e^{r,f}, \epsilon_B^{r,f}, E, \Gamma_{0}, \Delta_0),
\label{eq:para}
\end{equation}
where the superscript $r$ and $f$ represent reverse and forward shock respectively. 

\begin{figure*}[t]
\subfigure[]{
    \label{fig:subfig:a} %% label for first subfigure
    \includegraphics[width=3.0in]{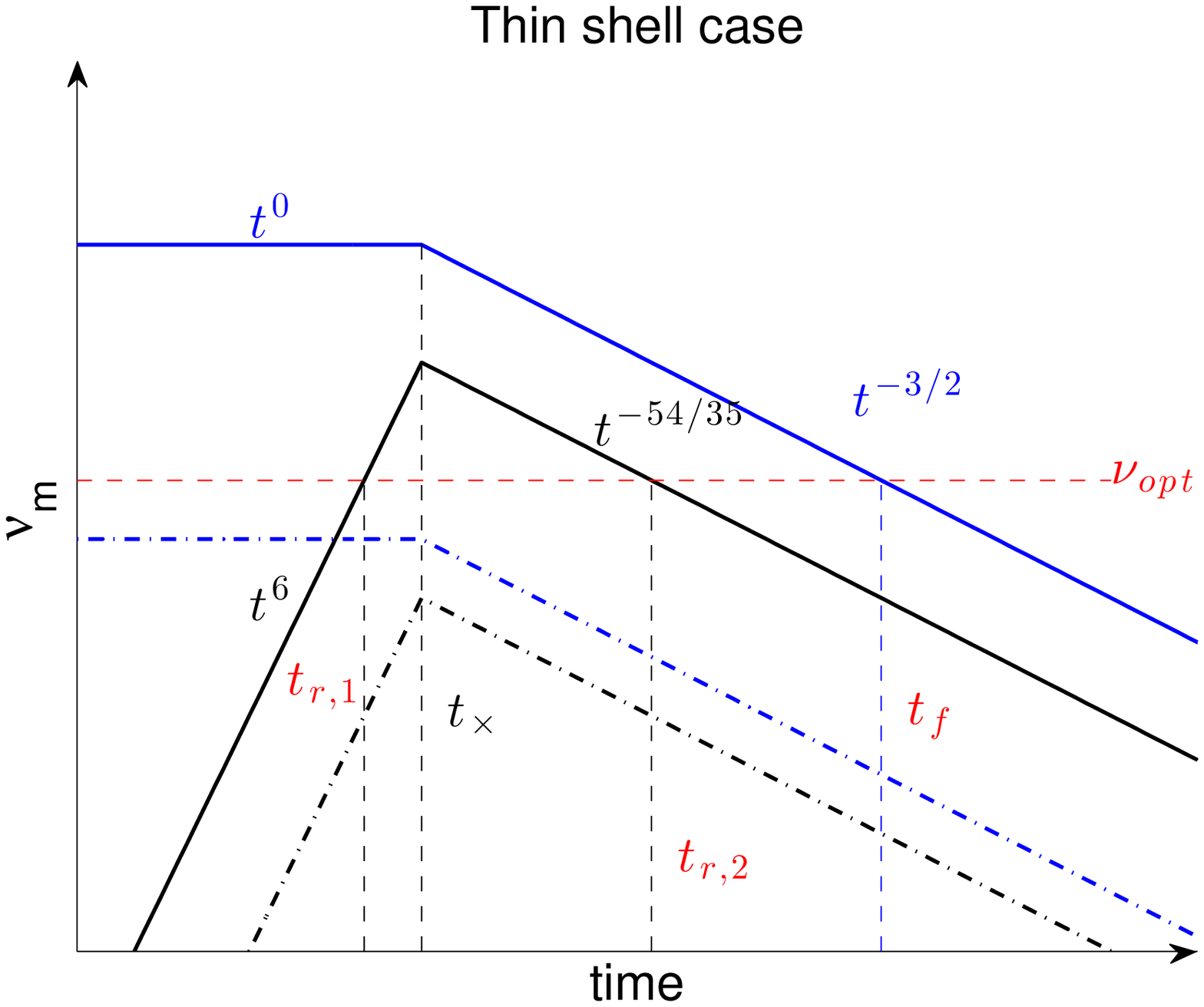}}
    \subfigure[]{
\label{fig:subfig:b} %% label for first subfigure
    \includegraphics[width=3.0in]{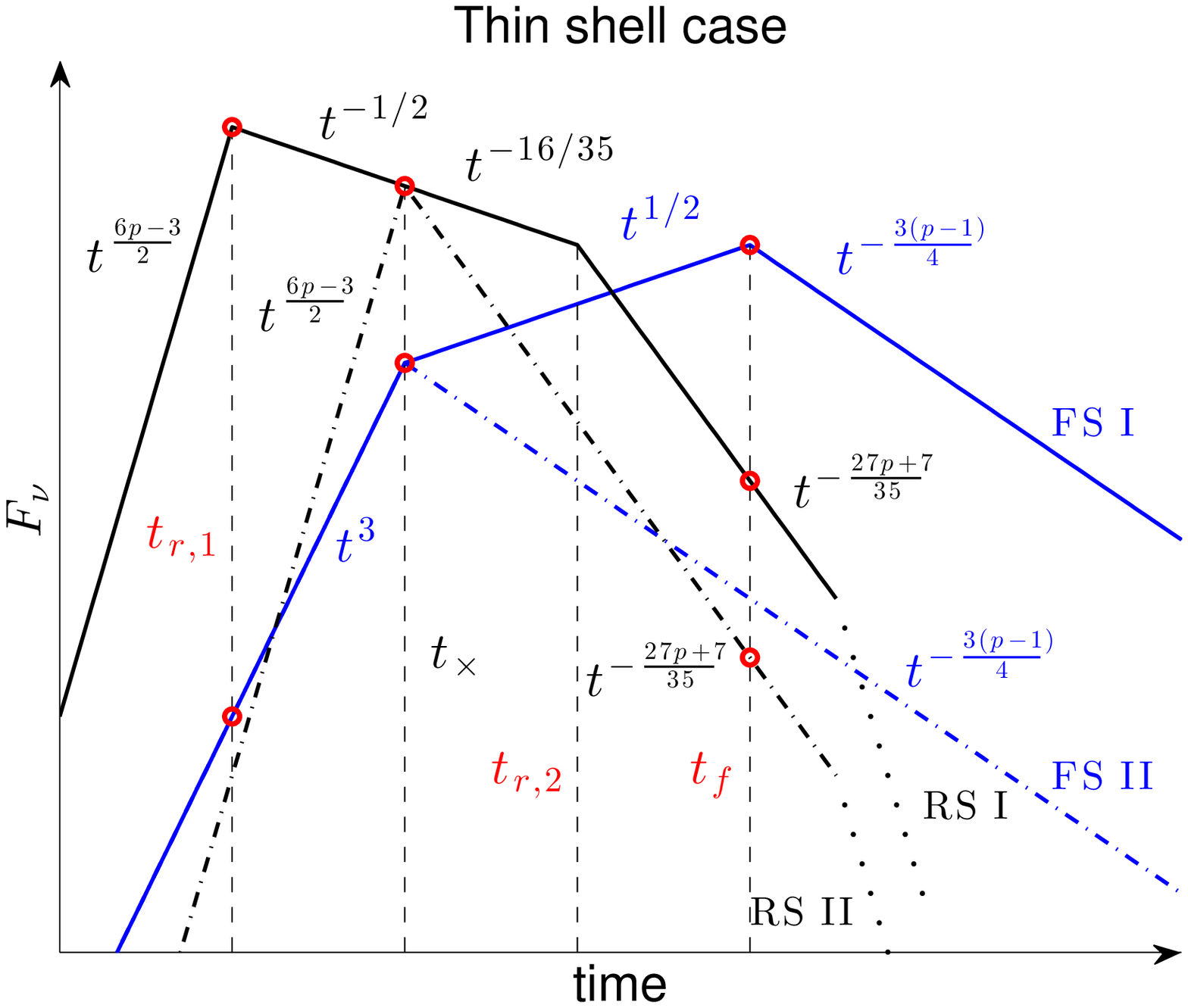}}\\
    \subfigure[]{
    \label{fig:subfig:c} %% label for first subfigure
    \includegraphics[width=3.0in]{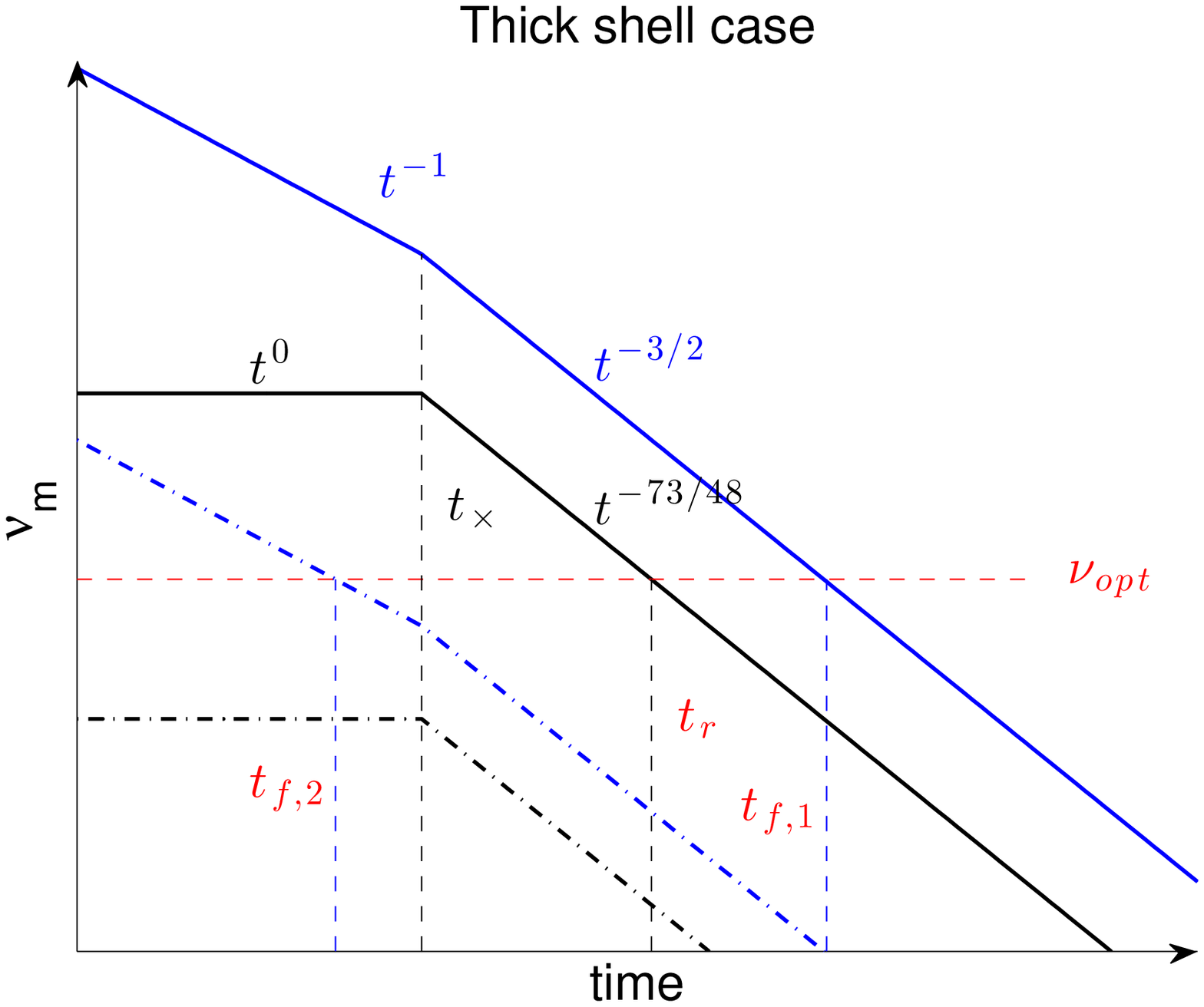}}
    \subfigure[]{
    \centering
    \label{fig:subfig:d} %% label for first subfigure
    \includegraphics[width=3.0in]{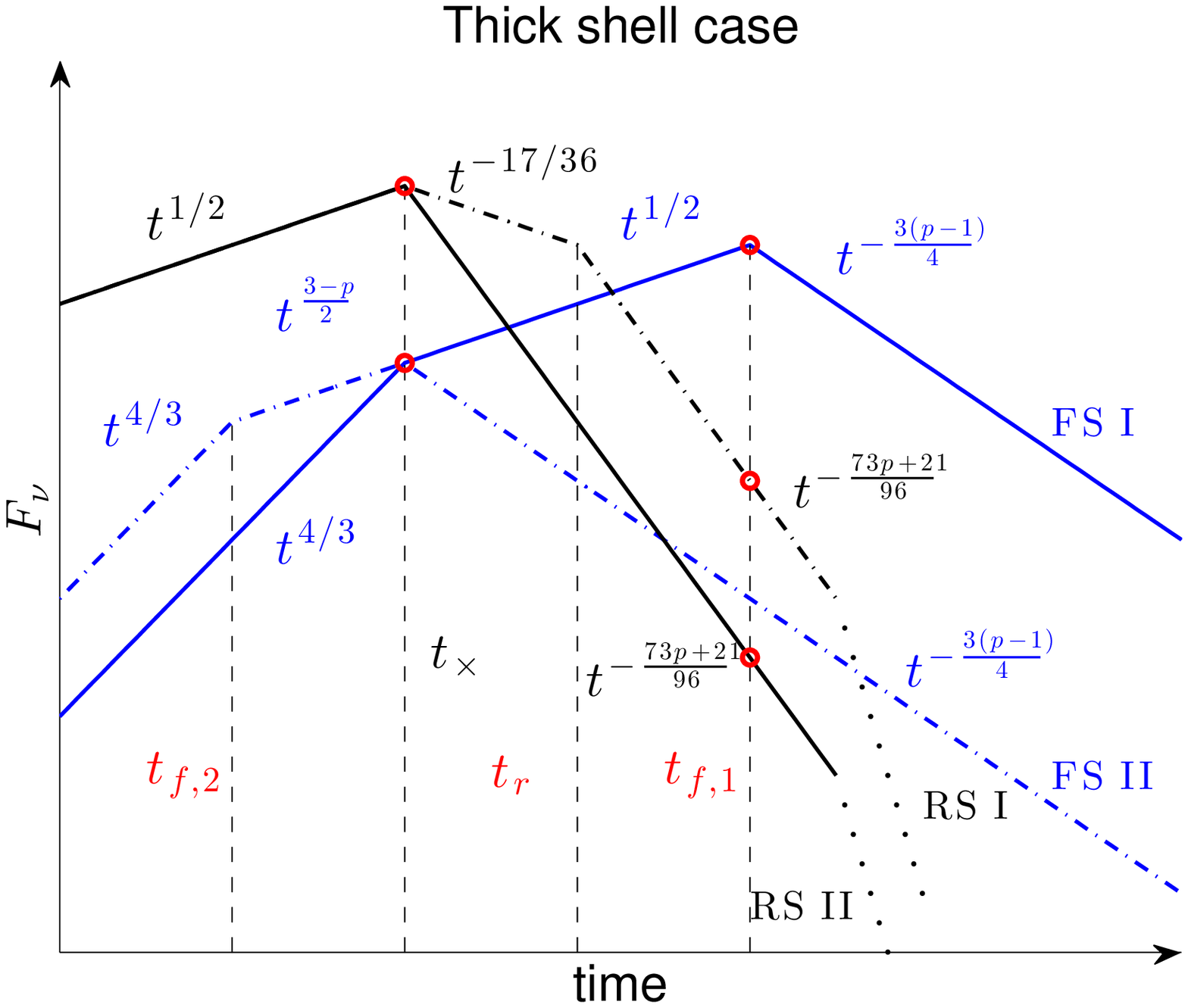}}\\
\caption{Illustration of the $\nu_m$ evolution (left panels) and optical lightcurves (right panels) 
for both forward shock (blue lines) and reverse shock (black lines) emission. Top panels are for thin 
shell regime and bottom panels are for thick shell regime. For  $\nu_m$ evolution panels, red solid 
line represents the observer frequency. For all panels, solid lines are for cases with 
$\nu^{r,f}_m(\tx)>\nuo$, dot-dash lines  are for cases with $\nu^{r,f}_m(\tx)<\nuo$. Red circles on 
lightcurve indicate the points for comparison in order to categorize the lightcurve types. Dotted 
lines at the end of reverse shock indicate the high latitude emission after $\nu_{c}^{r}$ crosses 
the observer frequency.}
\label{fig:illustration}          
\end{figure*}

In principle, the morphology for a specific GRB can be determined once the entire optical lightcurve 
is calculated. However, this process is time consuming and not conducive for realizing Monte Carlo 
simulation to explore a large parameter space. However, exploiting the power-law behavior of the
afterglow emission, here we propose an efficient scheme to categorize the lightcurve type, by 
comparing the reverse shock and forward shock flux strength at some special time point, rather than 
comparing them for the entire duration. The detailed scheme is illustrated as follows:

The dynamical evolution during the reverse shock crossing phase can be classified into two cases 
\citep{kobayashi00}, depending on whether the reverse shock becomes relativistic in the frame of  the
unshocked shell material (thick shell case) or not (thin shell case). Since the emissions from both 
reverse shock and forward shock behave differently in each case, our first step is to determine the 
thin/thick properties for given set of parameters. One practical way is to compare the duration of 
the burst $T=\Delta_0/c$ and the deceleration time of the ejecta $t_{\rm dec} = (3E/32\pi
nm_p\G_0^{8}c^{5})^{1/3}$, i.e., $T>t_{\rm dec}$ is for thick shell case and 
$T<t_{\rm dec}$ is for thin shell case, and $\tx= {\rm max} (t_{\rm dec}, T)$ \citep{zhang03}.

For the thin shell case, before shock crossing time, the evolution of 
$\nu_m^{r,f}$, $\nu_c^{r,f}$, and $F_{\rm{\nu,max}}^{r,f}$ reads\footnote{Since we focus on afterglow 
emission in optical band, the effect of $\nu_a$ is not considered here. } \citep{gao13review}
\begin{eqnarray}
&&\nu_m^r=1.9\times10^{12}~{\rm Hz}~\hat z^{-7}\frac{G(p)}{G(2.3)}E_{52}^{-2}\Gamma_{0,2}^{18}n_{0,0}
^{5/2}\epsilon_{e,-1}^{2}\epsilon_{B,-2}^{1/2}t_{2}^{6}           ,\nonumber\\
&&\nu_c^r=4.1\times10^{16}~{\rm Hz}~\hat z\Gamma_{0,2}^{-4}n_{0,0}^{-3/2}\epsilon_{B,-2}^{-3/2}t_{2}
^{-2}           \nonumber\\
&&F_{\rm{\nu,max}}^r=9.1\times10^{5}~\mu {\rm
Jy}~\hat z^{-1/2}E_{52}^{1/2} \Gamma_{0,2}^{5}n_{0,0}^{}\epsilon_{B,-2}^{1/2}D_{28}^{-2}t_{2}^{3/2}          
,\nonumber\\
&&\nu_m^f=   3.1\times10^{16}~{\rm Hz}~\hat z^{-1}\frac{G(p)}{G(2.3)}\Gamma_{0,2}^{4}n_{0,0}^{1/2}
\epsilon_{e,-1}^{2}\epsilon_{B,-2}^{1/2}           ,\nonumber\\
&&\nu_c^f=    4.1\times10^{16}~{\rm Hz}~\hat z\Gamma_{0,2}^{-4}n_{0,0}^{-3/2}\epsilon_{B,-2}^{-3/2}
t_{2}^{-2},           \nonumber\\
&&F_{\rm{\nu,max}}^f=     1.1\times10^{4}~\mu {\rm
Jy}~\hat z^{-2}\Gamma_{0,2}^{8}n_{0,0}^{3/2}\epsilon_{B,-2}^{1/2}D_{28}^{-2}t_{2}^{3}          ,\nonumber\\
\label{eq:thin}
\end{eqnarray}
where $G(p)=\left(\frac{p-2}{p-1}\right)^2$ and $\hat z=(1+z)$ is the redshift correction factor. For simplicity, we omit the superscript of $\epsilon_{e}^{r,f}$ and $\epsilon_{B}^{r,f}$. With the expression of $\tx$, the values of $\nu_m^{r,f}$, $\nu_c^{r,f}$, and $F_{\rm{\nu,max}}^{r,f}$ at the shock crossing time could be easily obtained. Consequently, with the temporal evolution power-law indices of these parameters \citep{gao13review}, one can calculate their values for the post shock crossing phase. At $\tx$, we have
\begin{eqnarray}
&&\frac{\nu_c^r(\tx)}{\nu_m^r(\tx)}=2.5\times10^{5}~\hat z^{8}E_{52}^{-2/3}\Gamma_{0,2}^{-2/3}n_{0,0}^{-4/3}\epsilon_{e,-1}^{-2}\epsilon_{B,-2}^{-2}           ,\nonumber\\
&&\frac{\nu_c^f(\tx)}{\nu_m^f(\tx)}=26.3~\hat z^{2}E_{52}^{-2/3}\Gamma_{0,2}^{-8/3}n_{0,0}^{-4/3}\epsilon_{e,-1}^{-2}\epsilon_{B,-2}^{-2}           .\nonumber\\
\label{eq:slowcooling}
\end{eqnarray}
where $G(p)$ factor is normalized to $p=2.3$. We can see that for the time we are interested in (e.g., mainly around or after $\tx$), both reverse shock and forward shock emission would be in the ``slow cooling'' regime ($\nu_{c}>\nu_{m}$) for reasonable parameter regimes \footnote{Note that for some extreme parameters, the fast cooling regime ($\nu_{c}<\nu_{m}$) might be relevant at the shock crossing time. However, in those cases, $\nu_c(\tx)$ could not be much smaller than $\nu_m(\tx)$, so that the real lightcurve shape would not deviate too much from the ones presented here (under the slow cooling assumption).} \citep{sari98}. We take slow cooling for both reverse and forward shock emission in the following, so that the shape of the light curve essentially depends on the relation between $\nu_{m}^{r,f}$ and $\nuo$ (similar arguments also apply to the thick shell case). The evolution of $\nu_{m}^{r,f}$ for the thin shell case is shown in Figure \ref{fig:illustration}a, which reads
\begin{eqnarray}
\nu_{m}^{f}\propto   t^{0}~(t<\tx),~\nu_{m}^{f}\propto   t^{-3/2}~(t>\tx),\nonumber\\
\nu_{m}^{r}\propto   t^{6}~(t<\tx),~\nu_{m}^{r}\propto   t^{-54/35}~(t>\tx).
\end{eqnarray}
When $\nu_{m}^{f}(\tx)$ is larger than $\nuo$, we call it FS I case (otherwise FS II case), and $\nu_{m}^{f}$ would cross the optical band once (at $t_{f}$). Similarly when $\nu_{m}^{r}(\tx)$ is larger than $\nuo$, we call it RS I case (otherwise RS II case), and $\nu_{m}^{r}$ would cross the optical band twice (at $t_{r,1}$ and $t_{r,2}$). There are altogether four combinations for different shapes of reverse and forward shock light curves. For each combination, we first check if the peak of the reverse shock emission is suppressed by the forward shock. If so, the lightcurve belongs to Type III (FS II) or IV (FS I). Otherwise, we will further check if the peak of the forward shock emission is suppressed by the reverse shock. If so, the lightcurve belongs to Type II, otherwise it is Type I. The evolution of $F_{\nu}^{r,f}$ for all four cases are presented in Table 1 and shown in Figure \ref{fig:illustration}b. The scheme to categorize the lightcurve type is presented in Table 2. 

\begin{table*}[tbp]
\begin{center}
\caption{The evolution of $F_{\nu}^{r,f}$ for all cases}
\begin{tabular}{c|c|c|c|c|c|c|c} \hline\hline
\multicolumn{5}{c|}{Thin Shell} & \multicolumn{3}{c}{Thick Shell}\\
  \hline
  \multicolumn{5}{c|}{FS I case ($t_f>t_{\times}$)} & \multicolumn{3}{c}{FS I case ($t_f>t_{\times}$)}\\
  \hline
 &\multicolumn{2}{c|}{$t<\tx$}                    & $\tx<t<t_f$   &$t>t_f$ &$t<\tx$                    & $\tx<t<t_f$   &$t>t_f$    \\
  \hline
$\Ff\propto$ &\multicolumn{2}{c|}{$t^{3}$}     &  $t^{1/2}$         &$t^{-3(p-1)/4}$ &  $t^{4/3}$     &  $t^{1/2}$         &$t^{-3(p-1)/4}$\\
   \hline
    \multicolumn{5}{c|}{FS II case ($t_f<t_{\times}$)} & \multicolumn{3}{c}{FS II case ($t_f<t_{\times}$)}\\
  \hline
 &\multicolumn{2}{c|}{$t<\tx$}                    & \multicolumn{2}{c|}{$t>\tx$} &$t<t_{f,2}$                    & $t_{f,2}<t<\tx$   &$t>\tx$    \\
  \hline
$\Ff\propto$ &\multicolumn{2}{c|}{$t^{3}$}     & \multicolumn{2}{c|}{$t^{-3(p-1)/4}$} &  $t^{4/3}$     &  $t^{(3-p)/2}$         &$t^{-3(p-1)/4}$\\
   \hline
    \multicolumn{5}{c|}{RS I case ($t_r>t_{\times}$)} & \multicolumn{3}{c}{RS I case ($t_r>t_{\times}$)}\\
  \hline
 &$t<t_{r,1}$                    & $t_{r,1}<t<\tx$   &$\tx<t<t_{r,2}$ &$t>t_{r,2}$ &$t<\tx$                    & $\tx<t<t_r$   &$t>t_r$    \\
  \hline
$\Fr\propto$ &$t^{(6p-3)/2}$     &  $t^{-1/2}$         &$t^{-16/35}$ &$t^{-(27p+7)/35}$ &  $t^{1/2}$     &  $t^{-17/36}$         &$t^{-(73p+21)/96}$\\
   \hline
    \multicolumn{5}{c|}{RS II case ($t_r<t_{\times}$)} & \multicolumn{3}{c}{RS II case ($t_r<t_{\times}$)}\\
  \hline
&\multicolumn{2}{c|}{$t<\tx$}                    & \multicolumn{2}{c|}{$t>\tx$} &$t<\tx$                     & \multicolumn{2}{c}{$t>\tx$}    \\
  \hline
$\Fr\propto$ &\multicolumn{2}{c|}{$t^{(6p-3)/2}$}     & \multicolumn{2}{c|}{$t^{-(27p+7)/35}$} &  $t^{1/2}$    & \multicolumn{2}{c}{$t^{-(73p+21)/96}$}\\
   \hline\hline
 \end{tabular}
\end{center}
\end{table*}

\begin{table*}[tbp]
\begin{center}{\scriptsize
\caption{The scheme to categorize the lightcurve type for all combinations for different shapes of reverse and forward shock light curves}
\begin{tabular}{c|c|c|c|c|c} \hline\hline
\multicolumn{3}{c|}{Thin Shell} & \multicolumn{3}{c}{Thick Shell}\\
  \hline
  \multicolumn{3}{c|}{FS I $+$ RS I}& \multicolumn{3}{c}{FS I $+$ RS I} \\
  \hline
$\Fr(t_{r,1})<\Ff(t_{r,1})$                   & \multicolumn{2}{c|}{$\Fr(t_{r,1})>\Ff(t_{r,1})$}    &$\Fr(\tx)<\Ff(\tx)$                   & \multicolumn{2}{c}{$\Fr(\tx)>\Ff(\tx)$}\\
  \hline
 &$\Ff(t_f)>\Fr(t_f)$     &  $\Ff(t_f)<\Fr(t_f)$         &  & $\Ff(t_{f,1})>\Fr(t_{f,1})$     &  $\Ff(t_{f,1})<\Fr(t_{f,1})$     \\
\cline{2-3}\cline{5-6}
  Type IV&  Type I   &  Type II        & Type IV&  Type I   &  Type II\\
   \hline
    \multicolumn{3}{c|}{FS I $+$ RS II}& \multicolumn{3}{c}{FS I $+$ RS II} \\
  \hline
$\Fr(\tx)<\Ff(\tx)$                   & \multicolumn{2}{c|}{$\Fr(\tx)>\Ff(\tx)$}    &$\Fr(\tx)<\Ff(\tx)$                   & \multicolumn{2}{c}{$\Fr(\tx)>\Ff(\tx)$}\\
  \hline
 &$\Ff(t_f)>\Fr(t_f)$     &  $\Ff(t_f)<\Fr(t_f)$         &  & $\Ff(t_{f,1})>\Fr(t_{f,1})$     &  $\Ff(t_{f,1})<\Fr(t_{f,1})$     \\
\cline{2-3}\cline{5-6}
   Type IV&  Type I   &  Type II        & Type IV& Type  I   & Type  II\\
   \hline
    \multicolumn{3}{c|}{FS II $+$ RS I}& \multicolumn{3}{c}{FS II $+$ RS I} \\
  \hline
$\Fr(t_{r,1})<\Ff(t_{r,1})$                   & \multicolumn{2}{c|}{$\Fr(t_{r,1})>\Ff(t_{r,1})$}    &$\Fr(\tx)<\Ff(\tx)$                   & \multicolumn{2}{c}{$\Fr(\tx)>\Ff(\tx)$}\\
  \hline
 &$\Ff(\tx)>\Fr(\tx)$     &  $\Ff(\tx)<\Fr(\tx)$         &  & $\Ff(\tx)>\Fr(\tx)$     &  $\Ff(\tx)<\Fr(\tx)$     \\
\cline{2-3}\cline{5-6}
  Type  III&  Type I   &  Type II        & Type III& Type  I   & Type  II\\
   \hline
    \multicolumn{3}{c|}{FS II $+$ RS II}& \multicolumn{3}{c}{FS II $+$ RS II} \\
  \hline
$\Fr(\tx)<\Ff(\tx)$                   & \multicolumn{2}{c|}{$\Fr(\tx)>\Ff(\tx)$}    &$\Fr(\tx)<\Ff(\tx)$                   & \multicolumn{2}{c}{$\Fr(\tx)>\Ff(\tx)$}\\
  \hline
 &$\Ff(\tx)>\Fr(\tx)$     &  $\Ff(\tx)<\Fr(\tx)$         &  & $\Ff(\tx)>\Fr(\tx)$     &  $\Ff(\tx)<\Fr(\tx)$     \\
\cline{2-3}\cline{5-6}
 Type   III& Type  I   & Type  II        & Type III&  Type I   &  Type II\\
   \hline\hline
 \end{tabular}
 }
\end{center}
\end{table*}

For the thick shell case, before shock crossing time $\tx$, the evolution of $\nu_m^{r,f}$, 
$\nu_c^{r,f}$, and $F_{\rm{\nu,max}}^{r,f}$ reads
\begin{eqnarray}
&&\nu_m^r=   7.6\times10^{11}~{\rm Hz}~\hat z^{-1}\frac{G(p)}{G(2.3)}\Gamma_{0,2}^{2}n_{0,0}^{1/2}\epsilon_{e,-1}^{2}\epsilon_{B,-2}^{1/2}           ,\nonumber\\
&&\nu_c^r=    1.2\times10^{17}~{\rm Hz}E_{52}^{-1/2}\Delta_{0,13}^{1/2}n_{0,0}^{-1}\epsilon_{B,-2}^{-3/2}t_{2}^{-1}           \nonumber\\
&&F_{\rm{\nu,max}}^r=     1.3\times10^{5} ~\mu {\rm
Jy}~\hat z^{1/2}E_{52}^{5/4} \Delta_{0,13}^{-5/4}\Gamma_{0,2}^{-1}n_{0,0}^{1/4}\epsilon_{B,-2}^{1/2}D_{28}^{-2}t_{2}^{1/2}          ,\nonumber\\
&&\nu_m^f=   1.0\times10^{16}~{\rm Hz}\frac{G(p)}{G(2.3)}E_{52}^{1/2}\Delta_{0,13}^{-1/2}\epsilon_{e,-1}^{2}\epsilon_{B,-2}^{1/2}t_{2}^{-1}           ,\nonumber\\
&&\nu_c^f=    1.2\times10^{17}~{\rm Hz}E_{52}^{-1/2}\Delta_{0,13}^{1/2}n_{0,0}^{-1}\epsilon_{B,-2}^{-3/2}t_{2}^{-1}           \nonumber\\
&&F_{\rm{\nu,max}}^f=     1.2\times10^{3}~\mu {\rm
Jy}~\hat z E_{52}^{} \Delta_{0,13}^{-1}n_{0,0}^{1/2}\epsilon_{B,-2}^{1/2}D_{28}^{-2}          .\nonumber\\
\end{eqnarray}
The evolution of $\nu_{m}^{r,f}$ for the thick shell case is shown in Figure \ref{fig:illustration}c, which reads
\begin{eqnarray}
\nu_{m}^{f}\propto   t^{-1}~(t<\tx),~\nu_{m}^{f}\propto   t^{-3/2}~(t>\tx),\nonumber\\
\nu_{m}^{r}\propto   t^{0}~(t<\tx),~\nu_{m}^{r}\propto   t^{-73/48}~(t>\tx).
\end{eqnarray}
Similar to the thin shell regime, we define four cases for different $F_{\nu}^{r,f}$ evolution, and present the results for all cases in Table 1 and Figure \ref{fig:illustration}d. The scheme to categorize the lightcurve type is presented in Table 2. 
Note that for thick shell case, Type III lightcurve could mimic like Type IV when $t_{f,2} \ll \tx$. 
However, the parameter space for this situation is very limited and practically this confusion could 
be easily clarified with spectral information, thus we still count this case as Type III in the 
simulation results. 

\begin{table*}[tbp]
\begin{center}{\scriptsize
\caption{Early optical afterglow properties for GRBs in the selected sample.}
\begin{tabular}{ccccccc} \hline\hline
GRB & $\alpha_{1}^a$ & $\alpha_{2}^b$ & $t_{b}^c$ & $f_{b}(\times 10^{-11})^d$ & Type & Data Reference\\
\hline
021004	&	1.50 	$\pm$	0.16 	&	1.06 	$\pm$	0.11 	&	100 			&	2.58			&	I	&	\cite{mirabal03}	\\
050525A	&	1.50 	$\pm$	0.20 	&	1.15 	$\pm$	0.10 	&	66 			&	23.23			&	I	&	\cite{blustin06}	\\
090424	&	1.51 	$\pm$	0.25 	&	0.85 	$\pm$	0.15 	&	176 			&	0.35			&	I	&	\cite{kann10}	\\
990123	&	2.52 	$\pm$	0.42 	&	1.52 	$\pm$	0.15 	&	42 	$\pm$	9 	&	698.64	$\pm$	140.00	&	II	&	\cite{castro-tirado99}	\\
021121	&	1.97 	$\pm$	0.15 	&	1.05 	$\pm$	0.11 	&	130 			&	0.20			&	II	&	\cite{liw03}	\\
050904	&	3.00 	$\pm$	0.27 	&	1.20 	$\pm$	0.12 	&	405 	$\pm$	40 	&	57.90	$\pm$	15.40	&	II	&	\cite{kann07}	\\
060111B	&	2.41 	$\pm$	0.21 	&	1.19 	$\pm$	0.11 	&	29 			&	6.16			&	II	&	\cite{stratta09}	\\
060117	&	2.42 	$\pm$	0.11 	&	1.00 	$\pm$	0.09 	&	109 			&	168.91			&	II	&	\cite{jelinek06}	\\
060908	&	1.48 	$\pm$	0.25 	&	1.05 	$\pm$	0.09 	&	50 			&	3.20			&	II	&	\cite{covino10}	\\
061126	&	2.00 	$\pm$	0.32 	&	0.86 	$\pm$	0.11 	&	23 			&	31.44			&	II	&	\cite{gomboc08}	\\
080319B	&	2.74 	$\pm$	0.42 	&	1.23 	$\pm$	0.15 	&	33 	$\pm$	12 	&	18246.50	$\pm$	7200.00	&	II	&	\cite{bloom09}	\\
081007	&	1.90 	$\pm$	0.21 	&	0.68 	$\pm$	0.08 	&	3 	$\pm$	1 	&	0.68	$\pm$	0.16	&	II	&	\cite{jin13}	\\
090102	&	1.84 	$\pm$	0.18 	&	1.10 	$\pm$	0.11 	&	50 	$\pm$	11 	&	8.07	$\pm$	3.00	&	II	&	\cite{gendre10}	\\
091024	&	2.20 	$\pm$	0.21 	&	1.05 	$\pm$	0.11 	&	430 	$\pm$	52 	&	3.50	$\pm$	1.10	&	II	&	\cite{virgili13}	\\
130427A	&	1.67 	$\pm$	0.11 	&	1.01 	$\pm$	0.07 	&	13 	$\pm$	3 	&	2359.09	$\pm$	500.00	&	II	&	\cite{vestrand14}	\\
030418	&	-0.81 	$\pm$	0.12 	&	0.55 	$\pm$	0.04 	&	1190 	$\pm$	109 	&	0.35	$\pm$	0.01	&	III	&	\cite{rykoff04}	\\
050820A	&	-2.00 	$\pm$	0.21 	&	1.10 	$\pm$	0.18 	&	500 	$\pm$	30 	&	2.00	$\pm$	0.30	&	III	&	\cite{cenko06}	\\
060110	&	-1.20 			&	0.80 			&	50 			&	15.00			&	III	&	\cite{liw06}	\\
060210	&	-1.19 	$\pm$	0.18 	&	1.33 	$\pm$	0.02 	&	718 	$\pm$	23 	&	0.10	$\pm$	0.01	&	III	&	\cite{curran07b}	\\
060418	&	-1.80 	$\pm$	0.10 	&	1.20 	$\pm$	0.03 	&	158 	$\pm$	1 	&	9.22	$\pm$	0.09	&	III	&	\cite{molinari07}	\\
060607A	&	-2.39 	$\pm$	0.18 	&	1.31 	$\pm$	0.11 	&	180 	$\pm$	15 	&	4.50	$\pm$	0.12	&	III	&	\cite{molinari07}	\\
060926	&	-4.02 	$\pm$	1.75 	&	0.75 	$\pm$	0.12 	&	78 	$\pm$	10 	&	0.17	$\pm$	0.03	&	III	&	\cite{lipunov06}	\\
061007	&	-2.99 	$\pm$	0.03 	&	1.67 	$\pm$	0.02 	&	90 	$\pm$	2 	&	179.09	$\pm$	0.86	&	III	&	\cite{mundell07b}	\\
070419A	&	-1.00 	$\pm$	0.12 	&	1.28 	$\pm$	0.03 	&	753 	$\pm$	32 	&	0.06	$\pm$	0.01	&	III	&	\cite{melandri09}	\\
070420	&	-1.43 	$\pm$	0.54 	&	0.90 	$\pm$	0.08 	&	196 	$\pm$	22 	&	1.80	$\pm$	0.14	&	III	&	\cite{klotz08}	\\
071010A	&	-1.06 	$\pm$	0.07 	&	0.72 	$\pm$	0.01 	&	384 	$\pm$	30 	&	0.46	$\pm$	0.02	&	III	&	\cite{covino08}	\\
071010B	&	-0.70 	$\pm$	0.37 	&	0.52 	$\pm$	0.03 	&	147 	$\pm$	36 	&	0.32	$\pm$	0.02	&	III	&	\cite{huangky09}	\\
071025	&	-1.20 	$\pm$	0.18 	&	1.11 	$\pm$	0.12 	&	563 	$\pm$	80 	&	0.07	$\pm$	0.01	&	III	&	\cite{perley10}	\\
071031	&	-0.74 	$\pm$	0.02 	&	0.76 	$\pm$	0.03 	&	1055 	$\pm$	11 	&	0.10	$\pm$	0.01	&	III	&	\cite{kruhler09b}	\\
080319A	&	-1.80 	$\pm$	0.08 	&	0.65 	$\pm$	0.07 	&	238 	$\pm$	17 	&	0.02	$\pm$	0.01	&	III	&	\cite{cenko08c}	\\
080603A	&	-3.85 	$\pm$	0.13 	&	1.17 	$\pm$	0.02 	&	482 	$\pm$	14 	&	1.04	$\pm$	0.05	&	III	&	\cite{guidorzi11}	\\
080710	&	-1.20 	$\pm$	0.12 	&	0.65 	$\pm$	0.07 	&	1695 	$\pm$	42 	&	0.47	$\pm$	1.10	&	III	&	\cite{kruhler09}	\\
080810	&	-1.15 	$\pm$	0.05 	&	1.14 	$\pm$	0.03 	&	142 	$\pm$	2 	&	14.28	$\pm$	0.21	&	III	&	\cite{page09}	\\
080928	&	-0.78 	$\pm$	0.05 	&	2.18 	$\pm$	0.19 	&	3223 	$\pm$	130 	&	0.40	$\pm$	0.01	&	III	&	\cite{rossi11}	\\
081008	&	-2.20 	$\pm$	0.09 	&	1.09 	$\pm$	0.04 	&	175 	$\pm$	1 	&	8.18	$\pm$	0.06	&	III	&	\cite{yuanf10}	\\
081126	&	-1.14 	$\pm$	0.02 	&	0.39 	$\pm$	0.01 	&	159 	$\pm$	2 	&	1.55	$\pm$	0.01	&	III	&	\cite{klotz09}	\\
081203A	&	-0.96 	$\pm$	0.03 	&	1.37 	$\pm$	0.02 	&	376 	$\pm$	1 	&	14.35	$\pm$	0.03	&	III	&	\cite{kuin09}	\\
090313	&	-1.36 	$\pm$	0.19 	&	0.92 	$\pm$	0.02 	&	1002 	$\pm$	69 	&	0.70	$\pm$	0.04	&	III	&	\cite{melandri10}	\\
090812	&	-1.35 	$\pm$	0.32 	&	1.37 	$\pm$	0.29 	&	71 	$\pm$	8 	&	1.79	$\pm$	0.11	&	III	&	\cite{wren09}	\\
091029	&	-3.10 	$\pm$	0.23 	&	0.48 	$\pm$	0.07 	&	312 	$\pm$	52 	&	0.14	$\pm$	1.00	&	III	&	\cite{marshall09}	\\
100219A	&	-1.50 	$\pm$	0.54 	&	0.95 	$\pm$	0.01 	&	619 	$\pm$	76 	&	0.15	$\pm$	0.02	&	III	&	\cite{maoj12}	\\
100901A	&	-1.87 	$\pm$	0.13 	&	1.00 	$\pm$	0.07 	&	1200 	$\pm$	95 	&	0.18	$\pm$	0.01	&	III	&	\cite{gorbovskoy12}	\\
100906A	&	-1.76 	$\pm$	0.04 	&	1.10 	$\pm$	0.03 	&	137 	$\pm$	1 	&	17.65	$\pm$	0.17	&	III	&	\cite{gorbovskoy12}	\\
110213A	&	-1.92 	$\pm$	0.02 	&	0.73 	$\pm$	0.04 	&	230 	$\pm$	1 	&	4.64	$\pm$	0.01	&	III	&	\cite{cucchiara11b}	\\
121217A	&	-1.80 	$\pm$	0.12 	&	0.80 	$\pm$	0.01 	&	1806 	$\pm$	29 	&	0.03	$\pm$	0.01	&	III	&	\cite{elliott14}	\\
060218	&	-0.35 	$\pm$	0.01 	&	0.78 	$\pm$	0.03 	&	55032 	$\pm$	1305 	&	0.03	$\pm$	0.01	&	IV	&	\cite{sollerman06}	\\
060605	&	-0.47 	$\pm$	0.05 	&	1.24 	$\pm$	0.01 	&	701 	$\pm$	38 	&	1.60	$\pm$	0.05	&	IV	&	\cite{rykoff09}	\\
070318	&	-0.59 	$\pm$	0.11 	&	1.07 	$\pm$	0.04 	&	456 	$\pm$	25 	&	1.64	$\pm$	0.04	&	IV	&	\cite{liang09}	\\
070411	&	-0.58 	$\pm$	0.14 	&	0.96 	$\pm$	0.01 	&	655 	$\pm$	25 	&	0.18	$\pm$	0.01	&	IV	&	\cite{ferrero08}	\\
080330	&	-0.25 	$\pm$	0.05 	&	0.97 	$\pm$	0.03 	&	902 	$\pm$	45 	&	0.21	$\pm$	0.01	&	IV	&	\cite{guidorzi09}	\\
090510	&	-0.53 	$\pm$	0.17 	&	1.13 	$\pm$	0.21 	&	1273 	$\pm$	501 	&	0.07	$\pm$	0.01	&	IV	&	\cite{pelassa10}	\\
120815A	&	-0.25 	$\pm$	0.03 	&	0.64 	$\pm$	0.01 	&	521 	$\pm$	21 	&	0.09	$\pm$	0.01	&	IV	&	\cite{kruhler13}	\\
120119A	&	-0.28 	$\pm$	0.08 	&	2.38 	$\pm$	0.90 	&	1021 	$\pm$	63 	&	0.27	$\pm$	0.01	&	IV	&	\cite{morgan14}	\\
060729	&	-1.20 	$\pm$	0.21 	&	0.90 	$\pm$	0.15 	&	800 	$\pm$	82 	&	0.60	$\pm$	0.02	&	h	&	\cite{grupe07}	\\
060904B	&	-0.95 	$\pm$	0.11 	&	1.07 	$\pm$	0.07 	&	493 	$\pm$	26 	&	0.40	$\pm$	0.01	&	h	&	\cite{klotz08}	\\
060906	&	-0.20 	$\pm$	0.45 	&	1.03 	$\pm$	0.35 	&	1263 	$\pm$	639 	&	0.05	$\pm$	0.01	&	h	&	\cite{rana09}	\\
070611	&	-2.58 	$\pm$	0.56 	&	0.90 	$\pm$	0.21 	&	2114 	$\pm$	342 	&	0.09	$\pm$	0.01	&	h	&	\cite{rykoff09}	\\
071112C	&	-0.60 	$\pm$	0.37 	&	0.91 	$\pm$	0.02 	&	165 	$\pm$	13 	&	0.32	$\pm$	0.02	&	h	&	\cite{huangky09}	\\
080319C	&	-0.38 	$\pm$	0.05 	&	2.15 	$\pm$	0.10 	&	654 	$\pm$	39 	&	0.21	$\pm$	0.01	&	h	&	\cite{liw08}	\\
081109A	&	-0.19 	$\pm$	0.18 	&	0.94 	$\pm$	0.03 	&	559 	$\pm$	128 	&	0.25	$\pm$	0.03	&	h	&	\cite{jin09}	\\
090726	&	-1.27 	$\pm$	0.12 	&	0.70 	$\pm$	0.18 	&	290 	$\pm$	45 	&	0.12	$\pm$	0.03	&	h	&	\cite{simon10}	\\
110205A	&	-3.54 	$\pm$	0.42 	&	1.51 	$\pm$	0.15 	&	958 	$\pm$	56 	&	3.24	$\pm$	0.33	&	h	&	\cite{cucchiara11}	\\
120711A	&	-0.50 	$\pm$	0.10 	&	1.03 	$\pm$	0.12 	&	332 	$\pm$	2 	&	7.15	$\pm$	1.05	&	h	&	\cite{martin-carrillo14}	\\
   \hline\hline
 \end{tabular}
 }
\end{center}
{\sc Note.} ---
\textbf{a:} For type I and II, $\alpha_{1}$ is the decaying slope of the reverse shock emission. For others, it represents the rising slope of the forward shock emission.
\textbf{b:} $\alpha_{2}$ is the decaying slope of the forward shock emission.
\textbf{c:} Peak time in unit of $s$. For type I and II, it is the peak time of reverse shock emission. For others, it is the peak time of forward shock emission.
\textbf{d:} Flux at peak time in unit of $\rm erg~ cm^{-2}~ s^{-1}$.
\end{table*}

\subsection{Sample Selection}
\label{sec:sample}

We systematically investigate all the Swift GRBs that have optical detections at times earlier than 
500 s after the prompt emission trigger, from the launch of Swift to March 2014. A sample of 114 
lightcurves is compiled either from published papers or from GCN Circulars if no published paper is 
available \citep{li12,liang13,kann10,kann11}. 

We first find the bursts without a detected initial rising. We fit their initial decaying phase with a single power-law function, and keep the bursts which have a decaying slope larger than 1.5 as candidates for Type I or Type II. Other bursts with relatively slower slopes are excluded from the following analysis since in principle they could belong to any one of the four types. 

Within the remaining sample, we find the bursts with rebrightening or flattening (steep decay to shallow decay) features. For these bursts, we fit their initial rising and decaying part with a smooth broken power-law function and take the bursts with decaying slope larger than 1.5 as candidates for Type I or Type II. All other bursts are taken as the candidate for Type III or Type IV.

For Type I/II candidates, we fit their lightcurves with two separate broken power-law components. If the peak flux for the weaker component is completely suppressed by the stronger component, the lightcurve is classified as Type II, otherwise it is classified as Type I. For Type III/IV candidates, we fit their lightcurves with one  broken power-law component, and take a rising slope smaller than 0.6 as the division between Type III and Type IV. For some bursts, there are early observations that can be used to exclude Type I and Type II, but it is hard to determine their rising slope to justify a Type III or Type IV classification in some cases, for instance, if the data points are too close to the peak or if the rising phase is superposed on an optical flare. We count these as an overlapping type in the following analysis. 

\begin{figure*}[t]
\subfigure[]{
    \label{fig:subfig:a} %% label for first subfigure
    \includegraphics[width=2.3in]{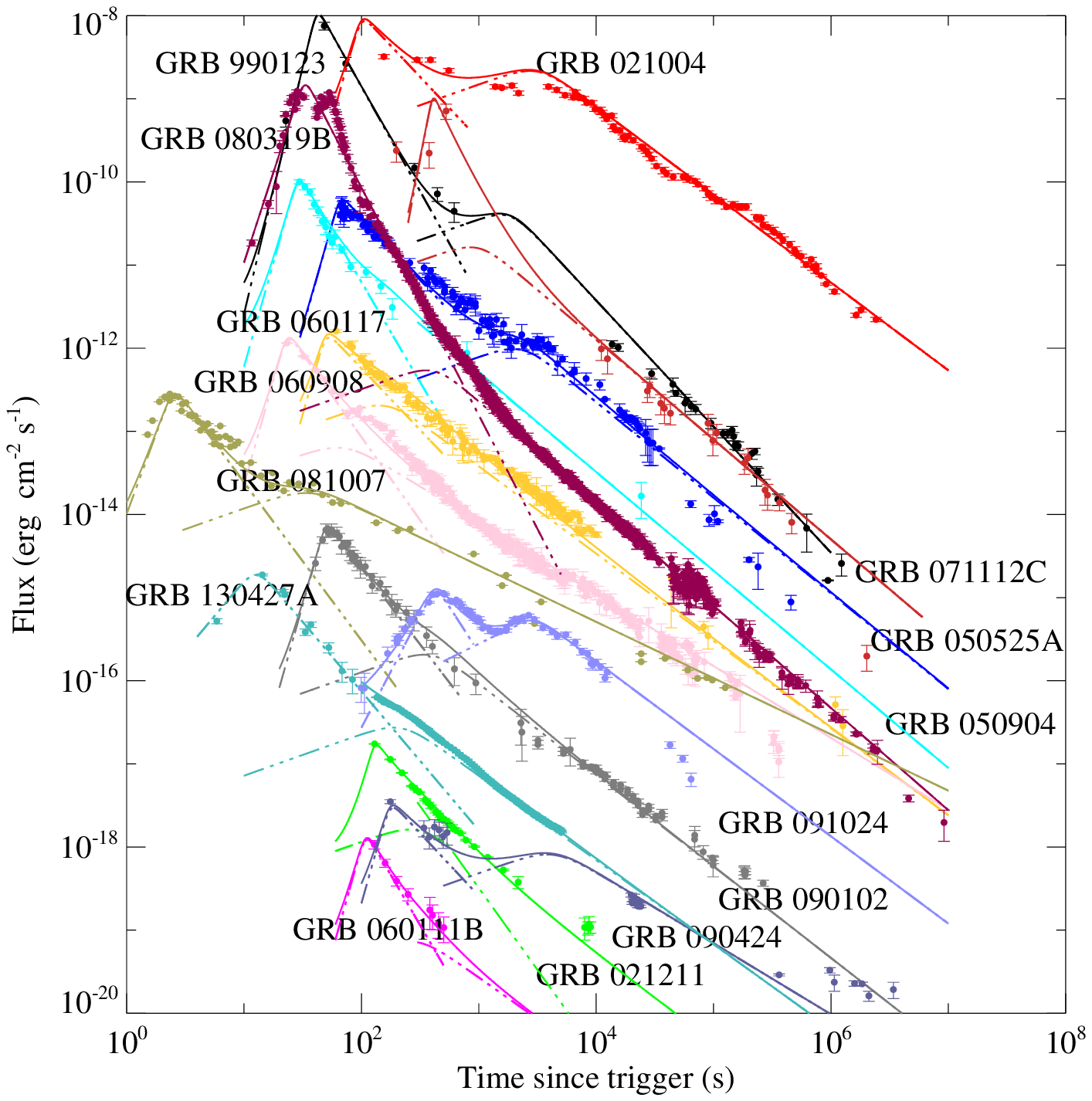}}
    \subfigure[]{
\label{fig:subfig:b} %% label for first subfigure
    \includegraphics[width=2.3in]{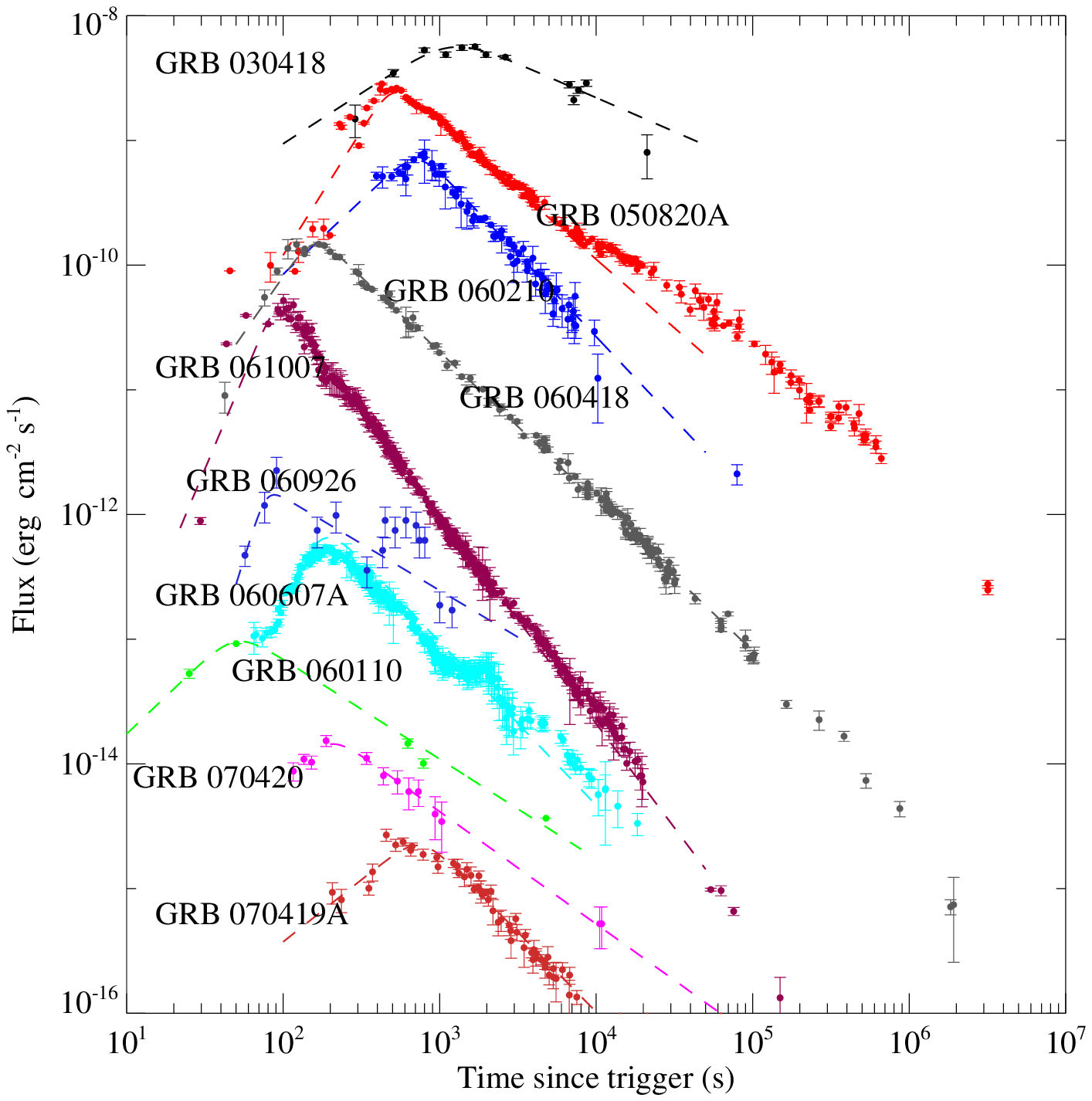}}
    \subfigure[]{
    \label{fig:subfig:c} %% label for first subfigure
    \includegraphics[width=2.3in]{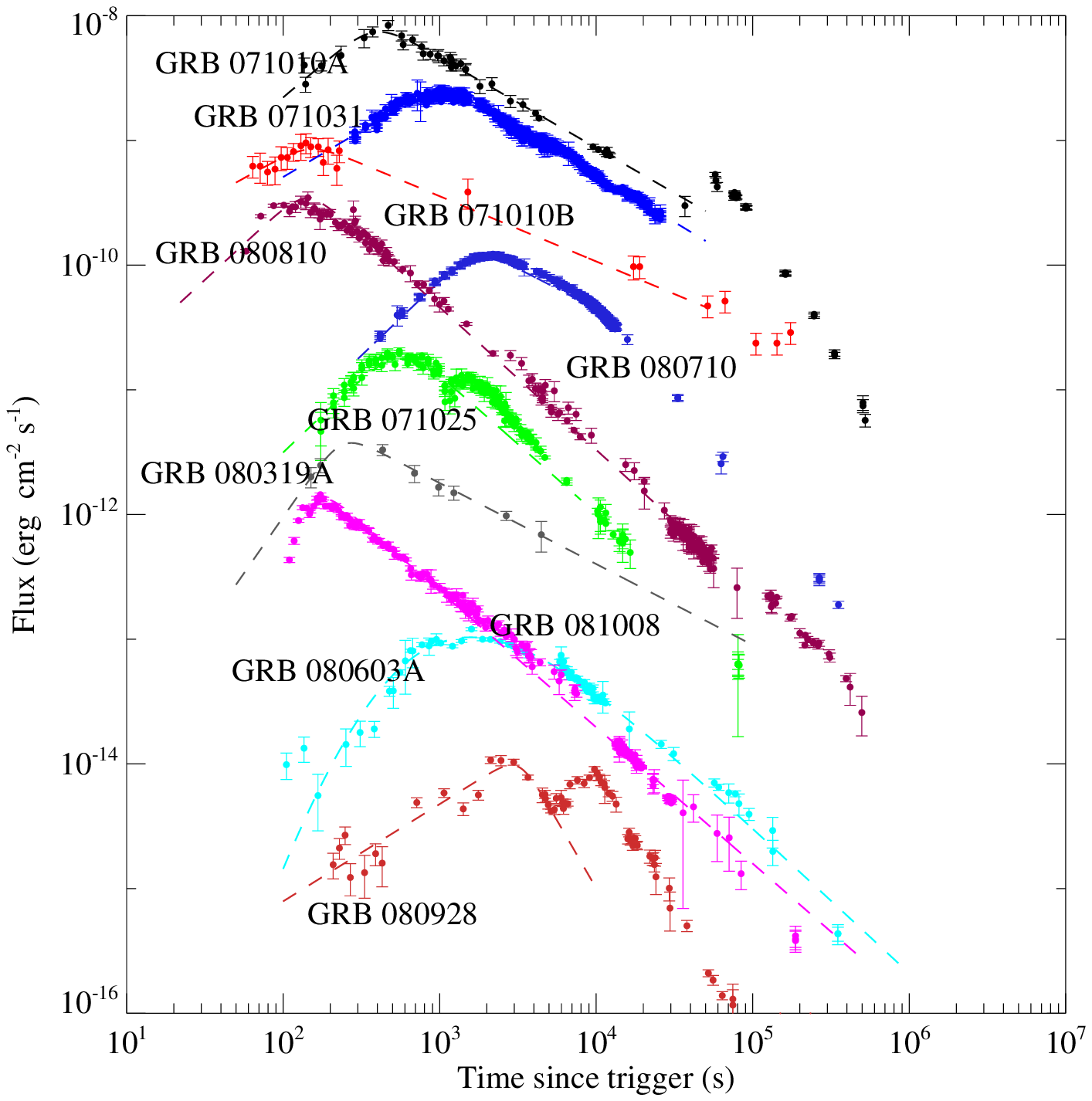}}\\
    \subfigure[]{
    \centering
    \label{fig:subfig:d} %% label for first subfigure
    \includegraphics[width=2.3in]{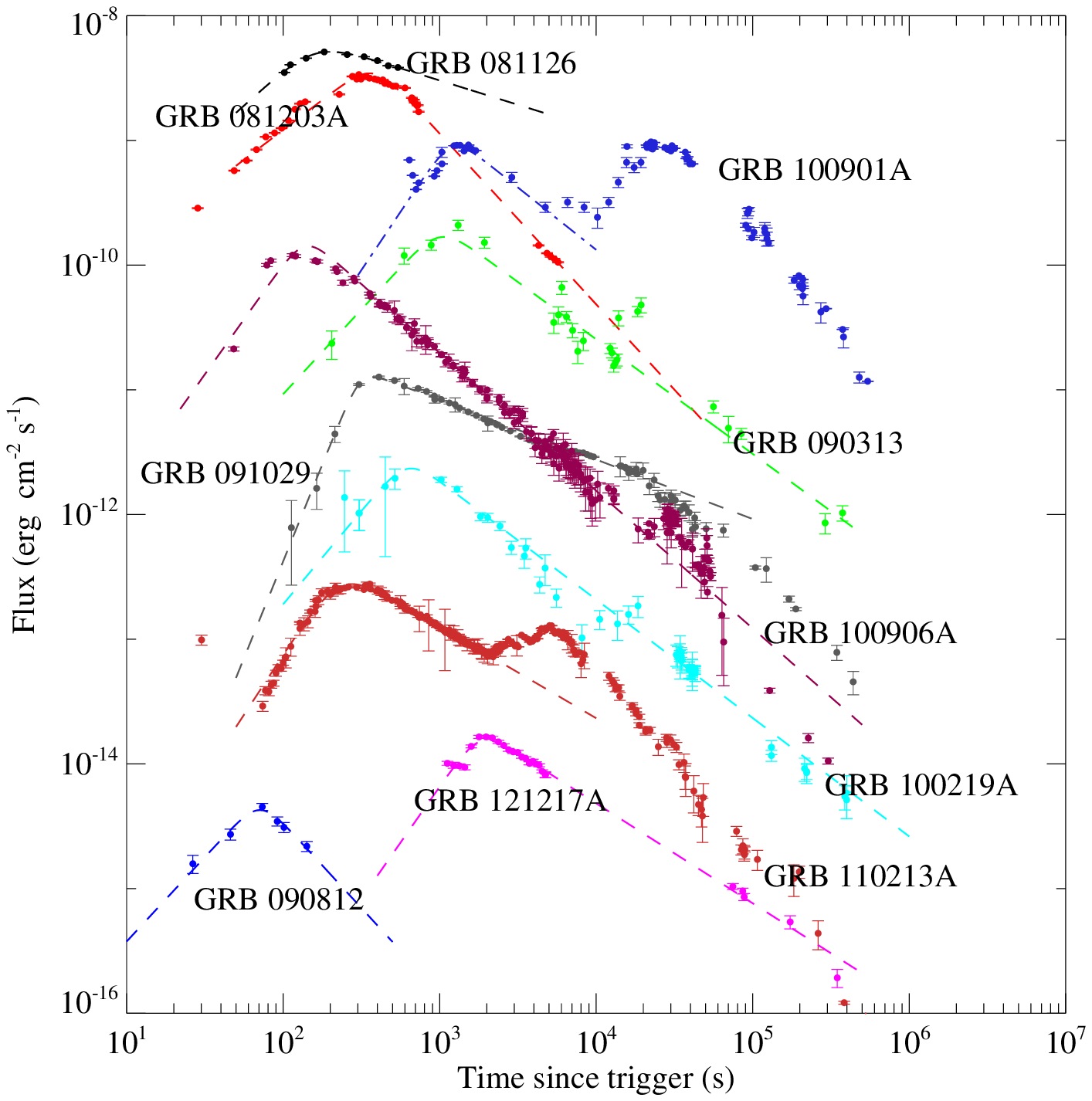}}
    \subfigure[]{
    \label{fig:subfig:e} %% label for first subfigure
    \includegraphics[width=2.3in]{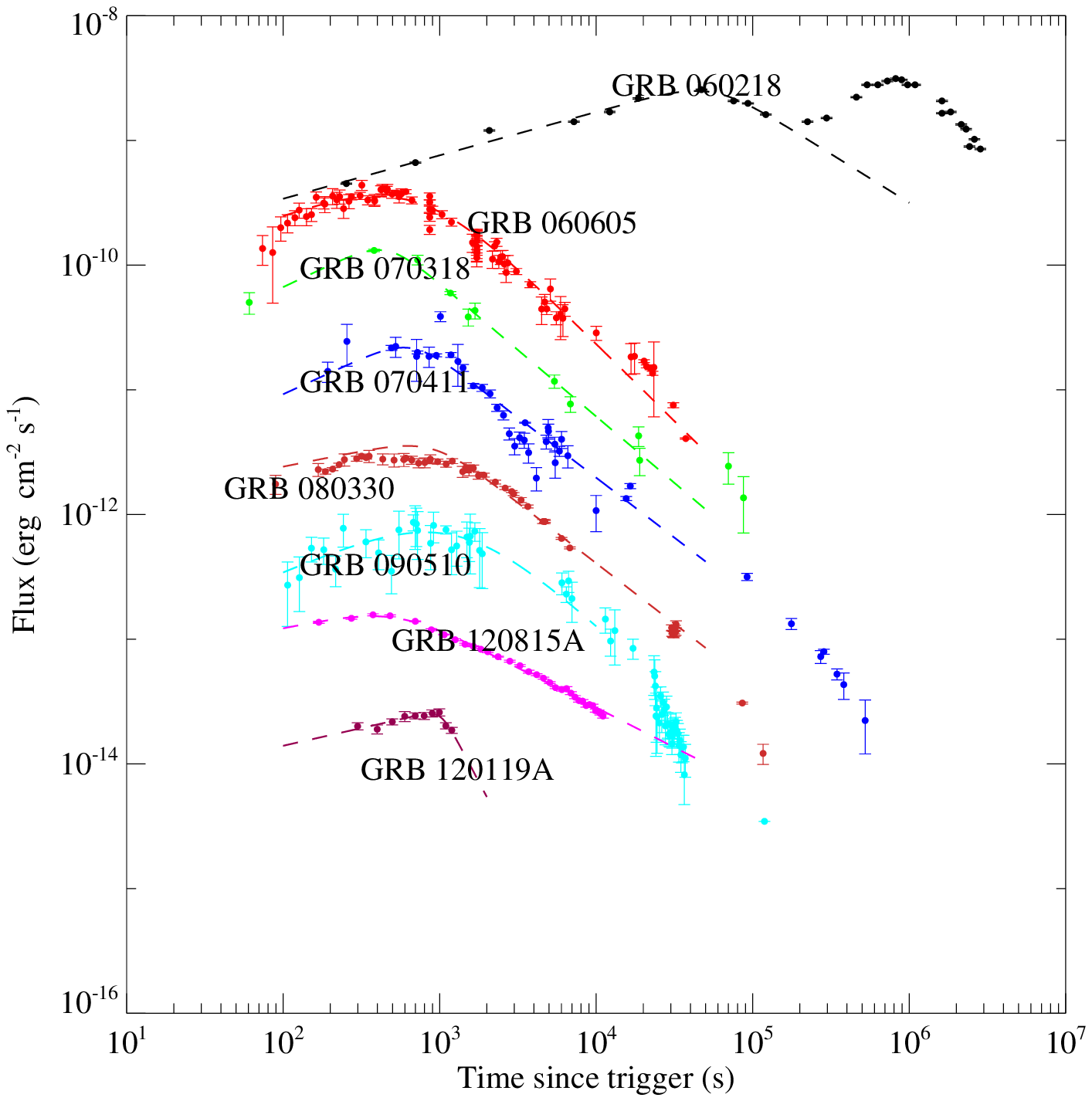}}
    \subfigure[]{
    \label{fig:subfig:e} %% label for first subfigure
    \includegraphics[width=2.3in]{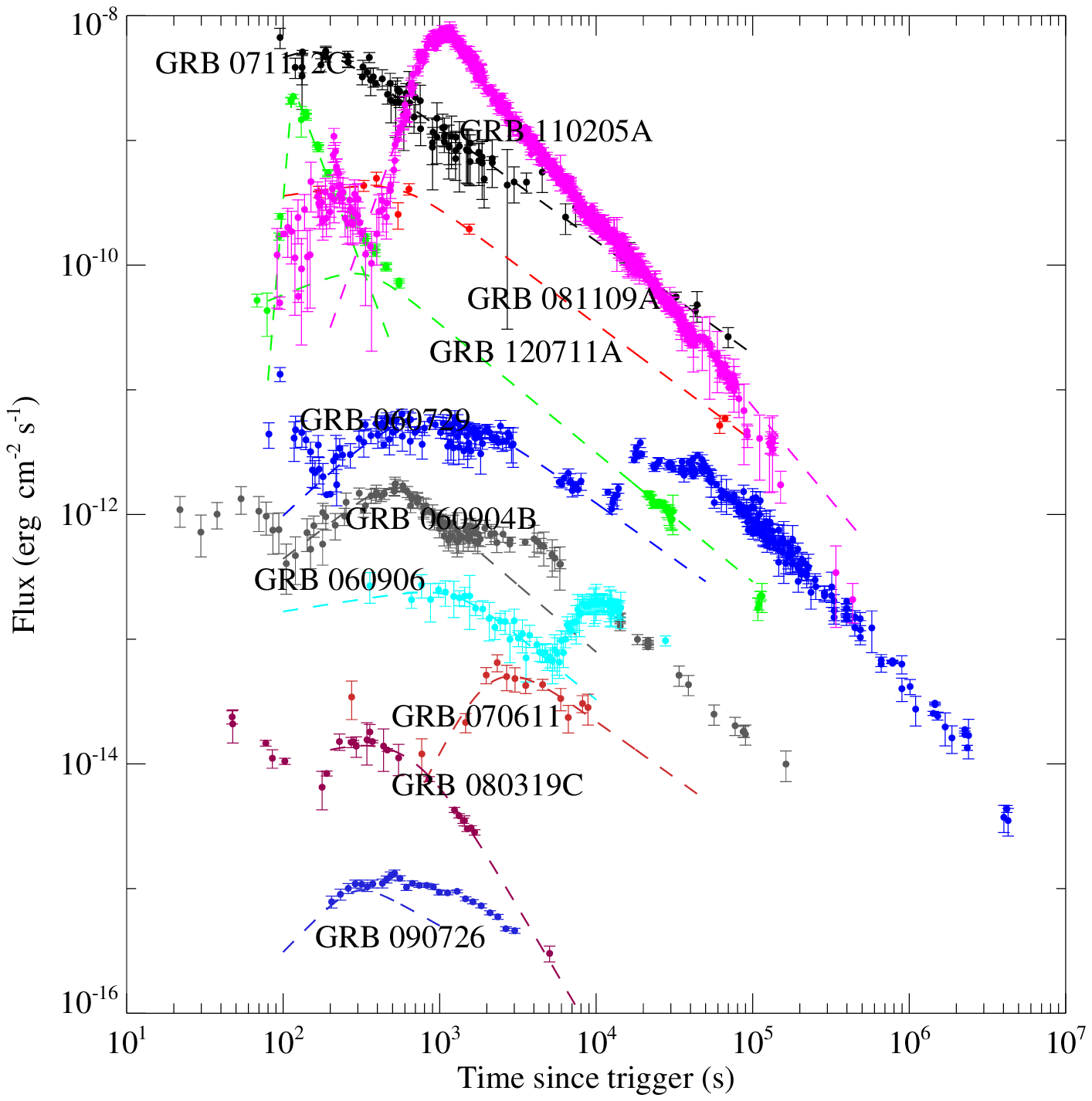}}
      \caption{Optical lightcurves and fitting results for different types in selected sample. (a) Type I and II; (b)-(d) Type III; (e) Type IV; (f) Type III/IV. The majority of the data are collected in terms of observed magnitudes. Since most data are in the R band, we first calibrate the data from other wavelengths (``X" band) to the R band with the expression $m_{R}=m_{X}-2.5 \beta_{O}\log_{10}(\lambda_{R}/\lambda_{X})+2.5\log_{10}(f_{0,R}/f_{0,X})$, where $\beta_{O}$ is the the optical spectral indices (assuming $F_{\nu}\propto \nu^{-\beta_{O}}$ being satisfied in optical band), and $f_{0}$ is the absolute spectral irradiance for $m=0.0$ within relevant magnitude system. An optical spectral index $\beta_{O}=0.75$ is adopted when $\beta_{O}$ is not available \citep{wang13,wang15}. We then convert the R band magnitudes to the flux in units of ${\rm erg~cm^{-2}~s^{-1}}$ with the expression $F_{R}=\lambda_{R}10^{(\log_{10}(f_{o,R})-0.4m_{R})}$, where $\lambda_{R}$ is the mean wavelength in R band. Galactic extinction correction is made to the data by using a reddening map presented by \cite{schlegel98}.}
           \label{fig:lightcurve}
            \end{figure*}

Eventually, we find 3 Type I bursts ($4.8\%$), 12 Type II bursts ($19.0\%$), 30 Type III bursts ($47.6\%$), 8 Type IV bursts ($12.7\%$) and 10 Type III/IV bursts ($15.9\%$). The lightcurves for each type are shown in Figure \ref{fig:lightcurve} and their properties are collected in Table 3, including the GRB name, the onset rising slope, decaying slope, peak time, peak flux and lightcurve type.

\section{Monte Carlo simulations}

In principle, if the intrinsic distribution function for each parameter listed in Equation \ref{eq:para} is available, we can simulate a sample of afterglow lightcurves, and distribute them into their relevant categories with the aforementioned theoretical scheme.  The properties of the parameter distributions could in turn be constrained by comparing the simulation results with the observational results collected in section \ref{sec:sample}. 

In the literature, several statistical works have been done through fitting individual bursts, either for late \citep{panaitescu01,panaitescu02,yost03} or early \citep{liang13,japelj14} broad-band observations. Although the intrinsic distribution functions are still poorly understood, some useful information, such as typical values or distribution ranges for most of the afterglow parameters, have been proposed \citep{kumarzhang15}. With this information, we could assume some proper distributions for the afterglow parameters, e.g. Gaussian distribution around the typical value or uniform distribution within a reasonable range. Even if the adopted distribution functions may be deviated from the intrinsic ones, it is still possible to justify how each parameter affects the result of morphological analysis. Nevertheless, for critical parameters that severely affect the results, their preferred values could be explored by comparing with the current observations.

\subsection{Simulation Setup}

The adopted distributions for generating afterglow parameters listed in Equation \ref{eq:para} are as 
follows: 

\begin{itemize}
\item The redshift $z$ is generated based on the assumption that the GRB rate roughly traces the star
formation history. We adopt a parameterized GRB rate model proposed by \cite{yuksel08}. 
\begin{equation}
R_{GRB}=\rho_{\rm0}\left[(1+z)^{-34}+\left(\frac{1+z}{5000}\right)^{3}+\left(\frac{1+z}{9}\right)^{35}\right]^{-0.1}.
\end{equation} 
The number of GRBs occurring per unit (observed) time in a comoving volume element $dV(z)/dz$ is then
\begin{equation}
\frac{dN}{dtdz}=\frac{R_{GRB}(z)}{1+z}\frac{dV(z)}{dz},
\end{equation}
where the $(1+z)$ factor accounts for the cosmological time dilation, and $dV(z)/dz$ is given by
\begin{equation}\label{volume}
\frac{dV(z)}{dz}=\frac{c}{H_{\rm 0}}\frac{4\pi D_{L}^2}{(1+z)^2
[\Omega_M(1+z)^3+\Omega_\Lambda]^{1/2}},
\end{equation}
for a flat $\Lambda$CDM universe. 
\item We assume that the electron spectral index $p$ is the same for reverse and forward shock. Recent investigations suggest that the distribution of $p$ is likely a Gaussian distribution, ranging from 2 to 3.5, with a typical value 2.5 \citep[e.g.][]{liang13,wang15}. To test the influence of p value on the final results, the distribution of p is taken to be a Gaussian distribution with standard deviation 0.2. Three values are tested for the mean value ($\bar{p}$), i.e., $2.3,~2.5,~2.7$.
\item The distribution function for the number density of the ISM medium is still with large uncertainty, roughly ranging from $0.1$ to $100~{\rm cm^{-3}}$ \citep{panaitescu01,panaitescu02}. Here we assume
a Gaussian distribution in log space for the ISM density. The standard deviation is fixed as $10^{0.6}$ (four orders of magnitude coverage for $3\sigma$) and three mean values ($\bar{n}$) are tested, i.e., 
$1,~10,~100~ \rm{cm^{-3}}$. 
\item We assume that the fractions of shock energy that go to electrons ($\epsilon_e^{r,f}$) are the same for the reverse and forward shocks. The distribution of $\epsilon_e^{r,f}$ is poorly constrained in the literature, but due to the energetics consideration, in the past, a convention value 0.1 has been assumed in most studies. Here we assume a Gaussian distribution in log space for $\epsilon_e^{r,f}$ with $10^{0.2}$ being the standard deviation\footnote{We also tested larger standard deviation values. It turns out that if the standard deviation for $\epsilon_e^{r,f}$ is too large, the observational results, especially the internal coordination between Type I and II could never be reproduced.}. Three values are tested for the mean value ($\bar{\epsilon}_{e}^{r,f}$), i.e., $0.001,~0.01,~0.1$.
\item The fractions of shock energy that go into magnetic fields ($\epsilon_B^{r,f}$) are very likely 
different, since the magnetization degree of the ejecta material tends to be larger than in the ISM 
medium. We define\footnote{This definition is different from the original definition of Zhang et al. (2003), who defined $\RB = B_r/B_f$, which is the square root of the $\RB$ defined in this paper.} 
\begin{eqnarray}
\RB \equiv\epsilon_B^{r}/\epsilon_B^{f}.
\end{eqnarray}
It has long been suggested that $\epsilon_B^{f}$ has a very wide distribution, ranging from $10^{-8}$ to $10^{-1}$ \citep{panaitescu01,santana14}. In the simulations, $\epsilon_B^{f}$ is 
generated through a  uniform distribution in log space with four ranges (each covering 3 orders of magnitude), i.e., $10^{-4}\sim10^{-1},~10^{-5}\sim10^{-2},~10^{-6}\sim10^{-3},~10^{-7}\sim10^{-4}$. $\RB$ is generated through a Gaussian distribution with mean values ($\RBbar$) $1,~10,~100$, and standard deviation $1,~\sqrt{10},~\sqrt{100}$ respectively. The value of $\epsilon_B^{r}$ could be calculated with $\epsilon_B^{f}$ and $\RB$ straightforwardly. 
\item Considering the power-law property of the luminosity function for GRBs \citep[e.g.][]{liang07}, the kinetic energy of the GRB ejecta is generated with a power-law distribution. The minimum and 
maximum energy value are fixed as $10^{50}~\rm{erg}$ and $10^{54}~\rm{erg}$ respectively \citep{zhang07a}. Three power-law index $\alpha_E$ are tested, i.e., $0.2,0.5,1$.
\item The initial Lorentz factor of the GRB ejecta is also suggested to have a wide range, from 50 to 500 \citep[e.g.][]{liang13}. Here we generate $\G_0$ with a uniform distribution in the log space. We test three combinations of the minimum and maximum values, i.e., $50\sim300$, $100\sim500$ and $50\sim500$.
\item The observed shell width essentially shares the same distribution with the GRB duration, which could be well described with a Gaussian distribution in the log space\footnote{We only consider long GRBs here, since the collected sample are essentially all long GRBs. } \citep[e.g.][]{qin13}. The observed shell width is generated with a Gaussian distribution in log space with standard deviation $10^{0.6}$. We test two mean values ($\bar{\Delta}_0$), i.e., $10^{11}$ cm and $10^{12}$ cm.
\end{itemize}

\subsection{Simulation Results}

Given a set of distribution functions for each afterglow parameter, we run Monte Carlo simulation for 
10000 times \footnote{The number of runs for each simulation is determined by balancing the computation 
time consumption and the resulting convergence.}, and we analyze the thus obtained distributions of 
fractional ratios between different types of lightcurves. In Figures \ref{fig:simu1}-\ref{fig:simu2} 
we plot the simulation results for selected situations which are relevant for illustrating the main 
conclusions, which can be summarized as follows:

\begin{figure*}[t]
\subfigure[$\bar{\epsilon}_{e}^{f}=0.1,\RBbar=1$]{
    \label{fig:subfig:a} %% label for first subfigure
    \includegraphics[width=2.0in]{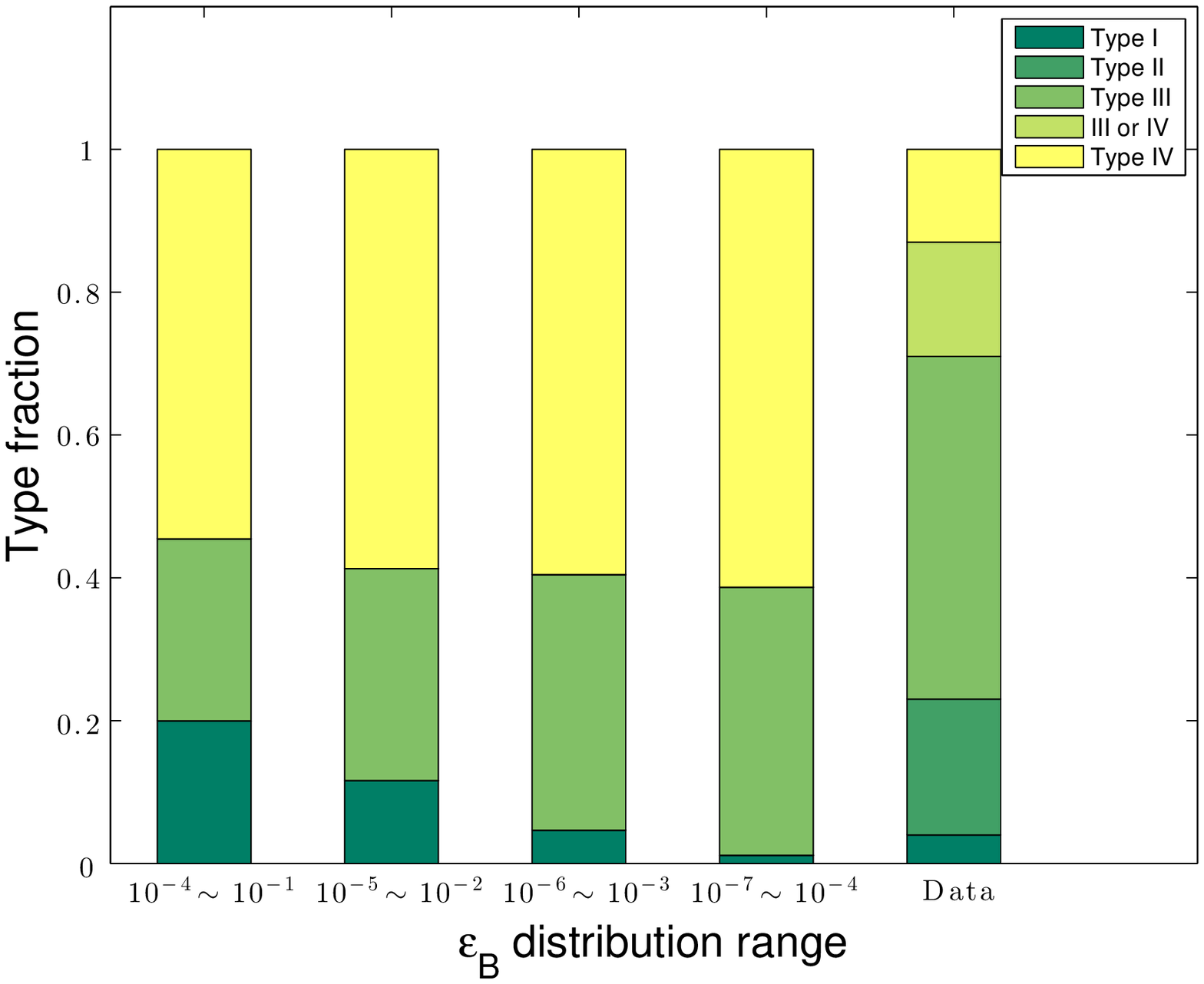}}
    \subfigure[$\bar{\epsilon}_{e}^{f}=0.1,\RBbar=10$]{
\label{fig:subfig:b} %% label for first subfigure
    \includegraphics[width=2.0in]{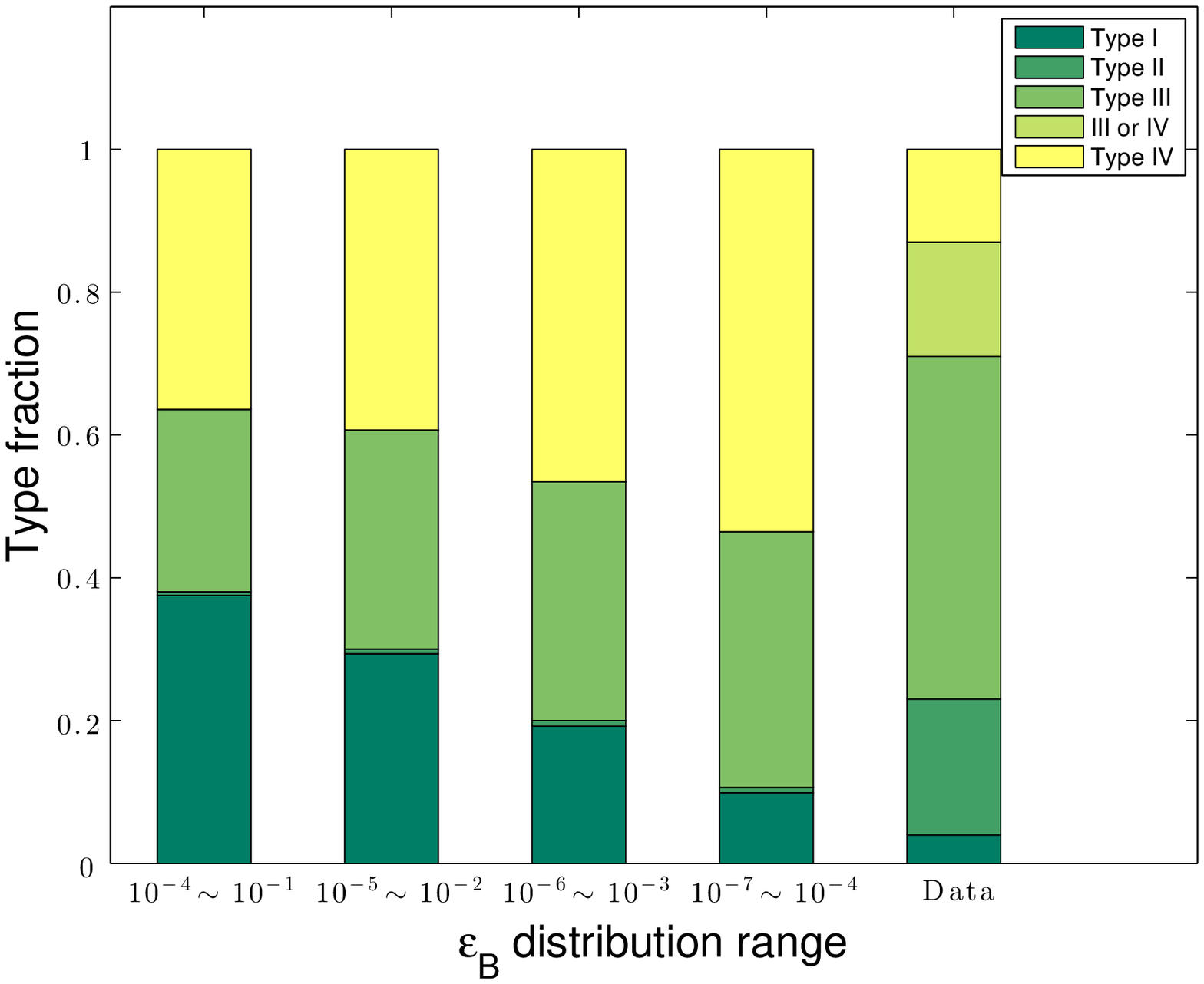}}
    \subfigure[$\bar{\epsilon}_{e}^{f}=0.1,\RBbar=100$]{
    \label{fig:subfig:c}%% label for first subfigure
    \includegraphics[width=2.0in]{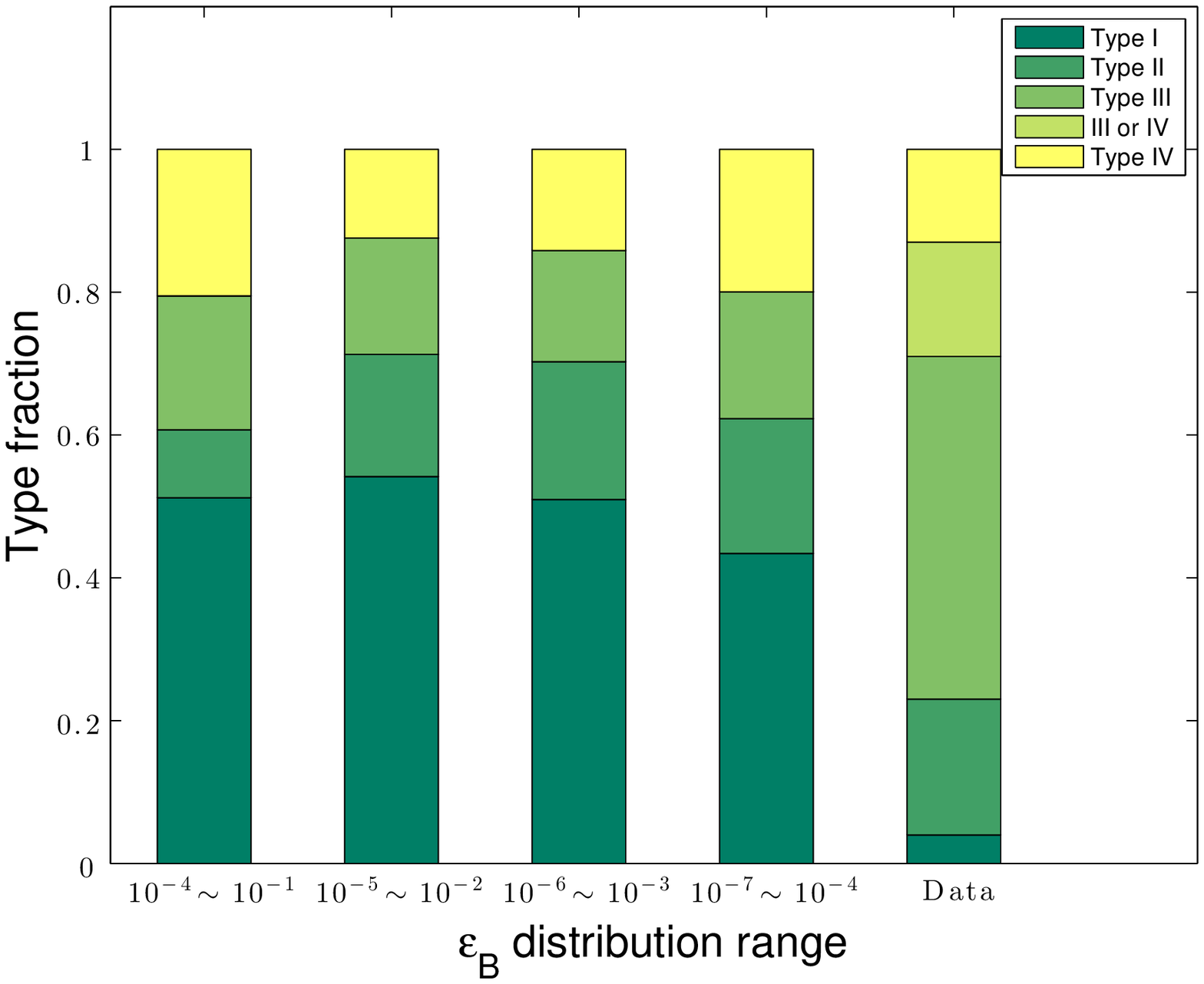}}\\
    \subfigure[$\epsilon_e^{f}=0.1,\RB=10,\bar{n}=10$]{
    \centering
    \label{fig:subfig:d} %% label for first subfigure
    \includegraphics[width=2.0in]{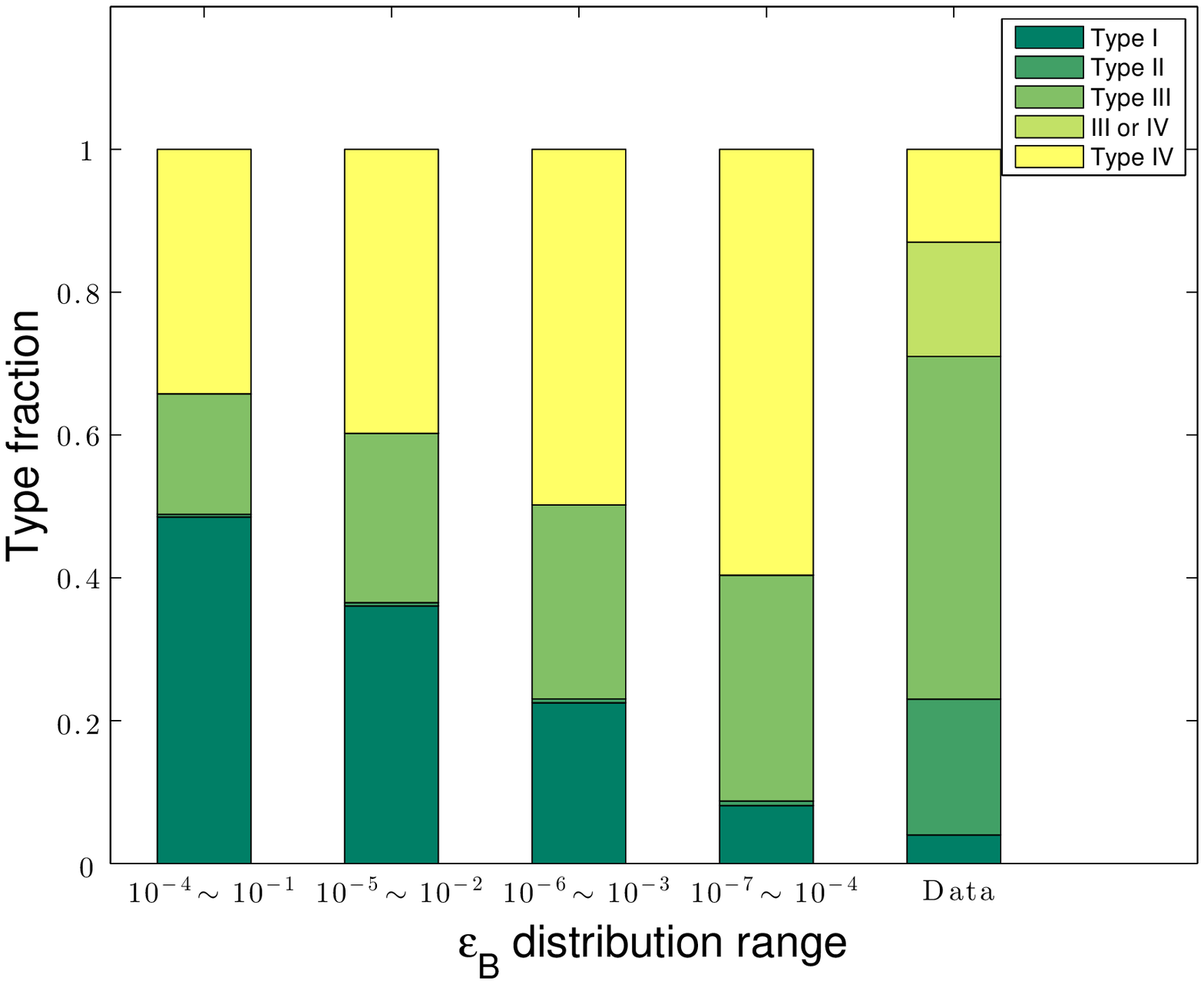}}
    \subfigure[$\epsilon_e^{f}=0.1,\RB=10,\bar{n}=100$]{
    \label{fig:subfig:a}%% label for first subfigure
    \includegraphics[width=2.0in]{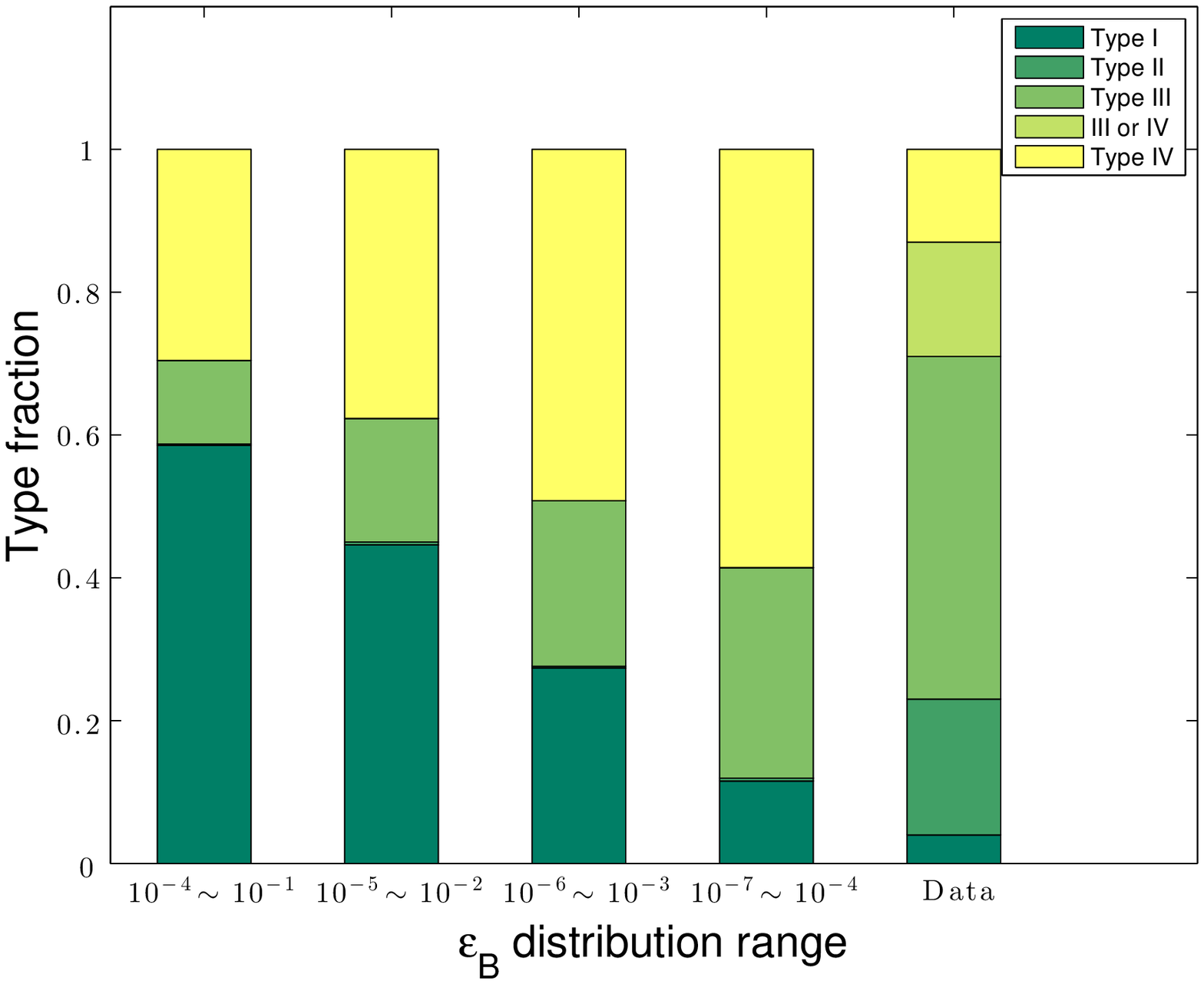}}
     \subfigure[$\epsilon_e^{f}=0.1,\RB=10,\bar{p}=2.5$]{
    \label{fig:subfig:a}%% label for first subfigure
    \includegraphics[width=2.0in]{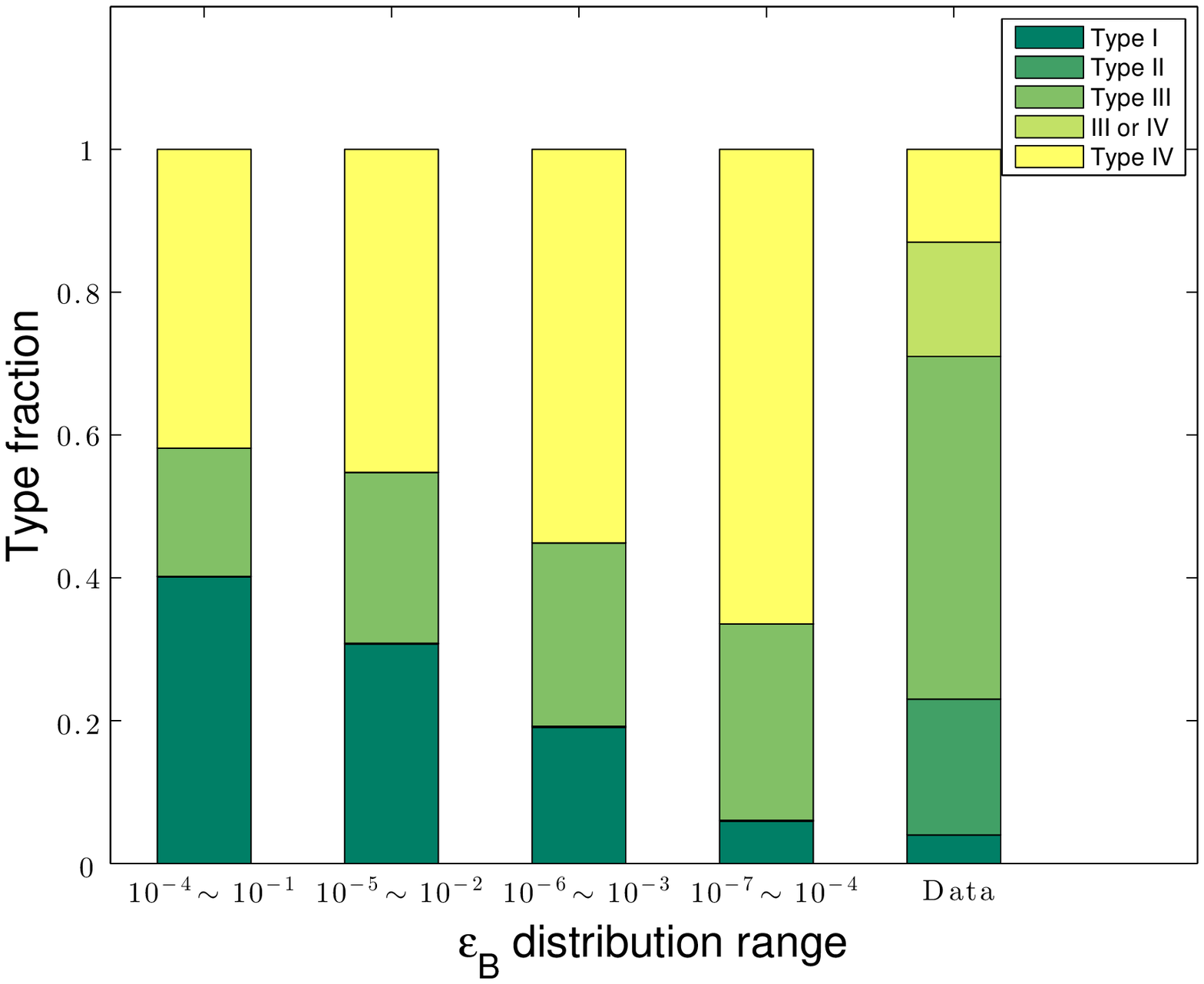}}\\
    \subfigure[$\epsilon_e^{f}=0.1,\RB=10,\bar{p}=2.7$]{
    \label{fig:subfig:a} %% label for first subfigure
    \includegraphics[width=2.0in]{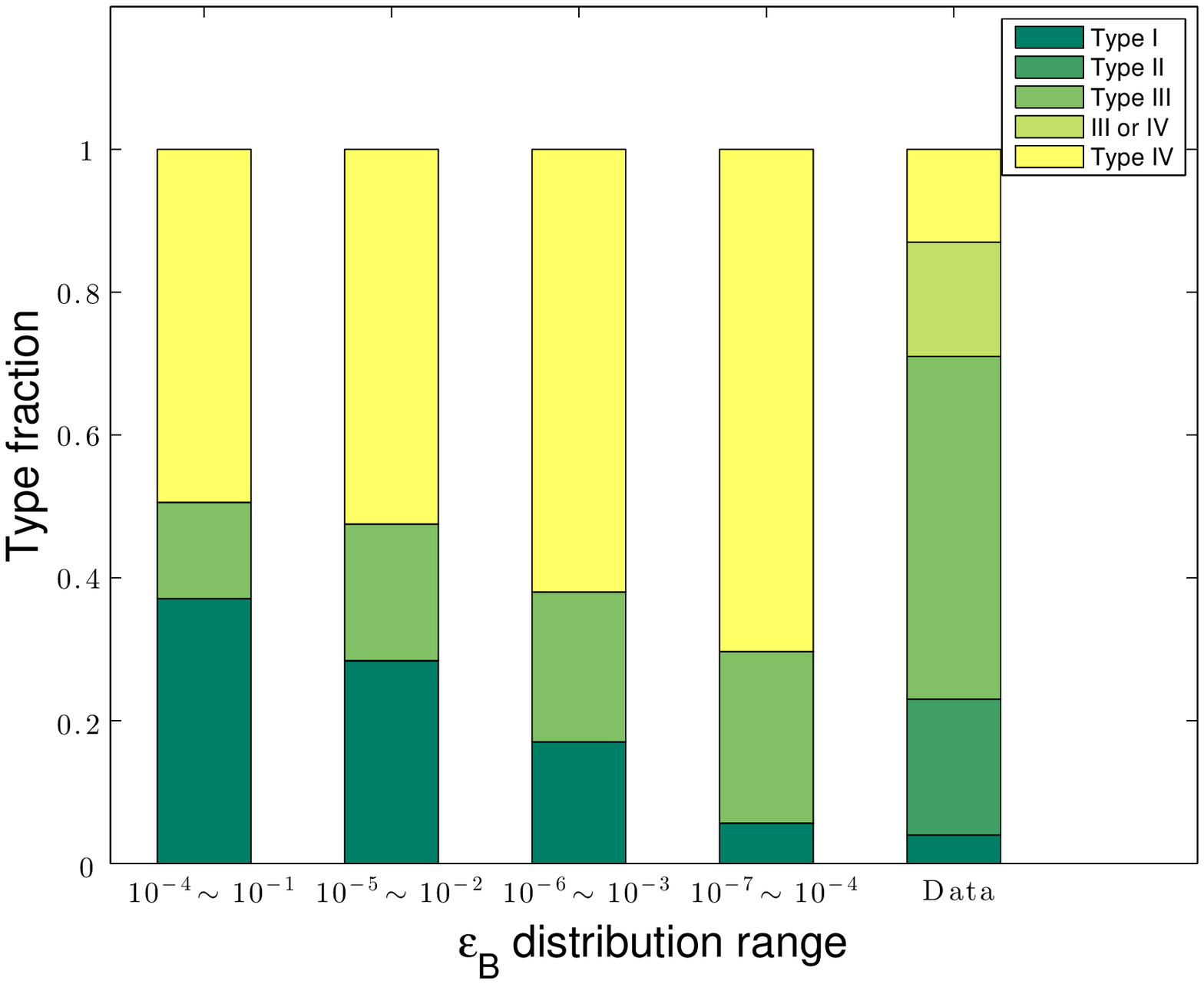}}
    \subfigure[$\epsilon_e^{f}=0.1,\RB=10,\G_0=100\sim500$]{
    \label{fig:subfig:a}%% label for first subfigure
    \includegraphics[width=2.0in]{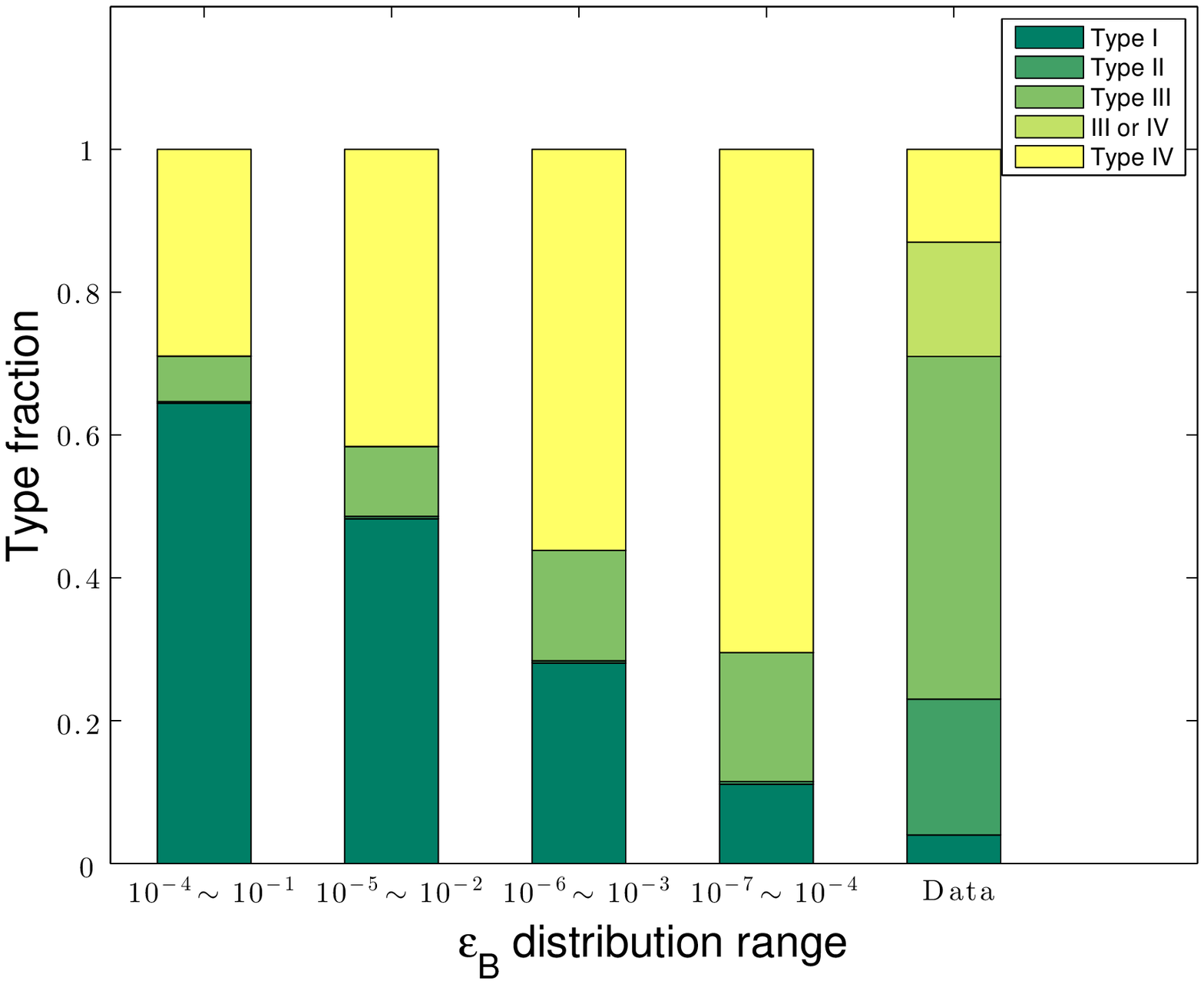}}
    \subfigure[$\epsilon_e^{f}=0.1,\RB=10,\G_0=50\sim500$]{
    \label{fig:subfig:a}%% label for first subfigure
    \includegraphics[width=2.0in]{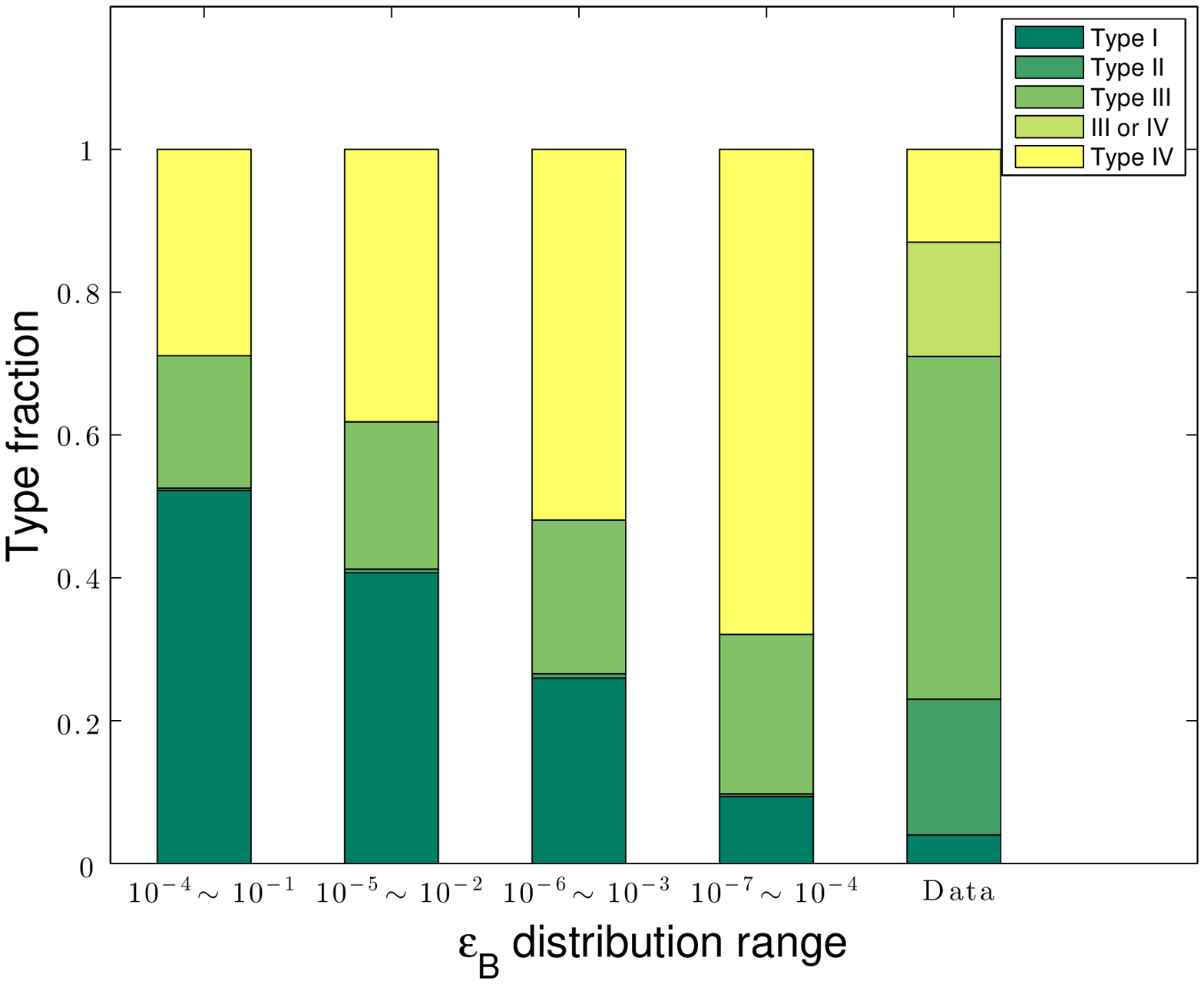}}\\
\caption{Stacked fractions of different lightcurve types for observational data and selected simulation 
results. Each panel corresponds to one specific simulation setup. Unless specified under the subfigure, 
the general setup values for different parameters are as follows (see details in section 3.1): 
$\bar{n}=1,\bar{p}=2.3,\G_0=50\sim300,\alpha_E=-0.5,\bar{\Delta}_0=10^{11}$. Five stacked histograms 
are presented in each panel. The first four histograms are for different $\epsilon_B^{f}$ ranges 
(left to right: $10^{-4}\sim10^{-1},~10^{-5}\sim10^{-2},~10^{-6}\sim10^{-3} {\rm and}~10^{-7}\sim10^{-4}$)
and the last one represents the observational results. }
           \label{fig:simu1}
            \end{figure*}

\begin{figure*}[t]
 \subfigure[$\epsilon_e^{f}=0.1,\RB=10,\alpha_E=-0.2$]{
    \label{fig:subfig:a}%% label for first subfigure
    \includegraphics[width=2.0in]{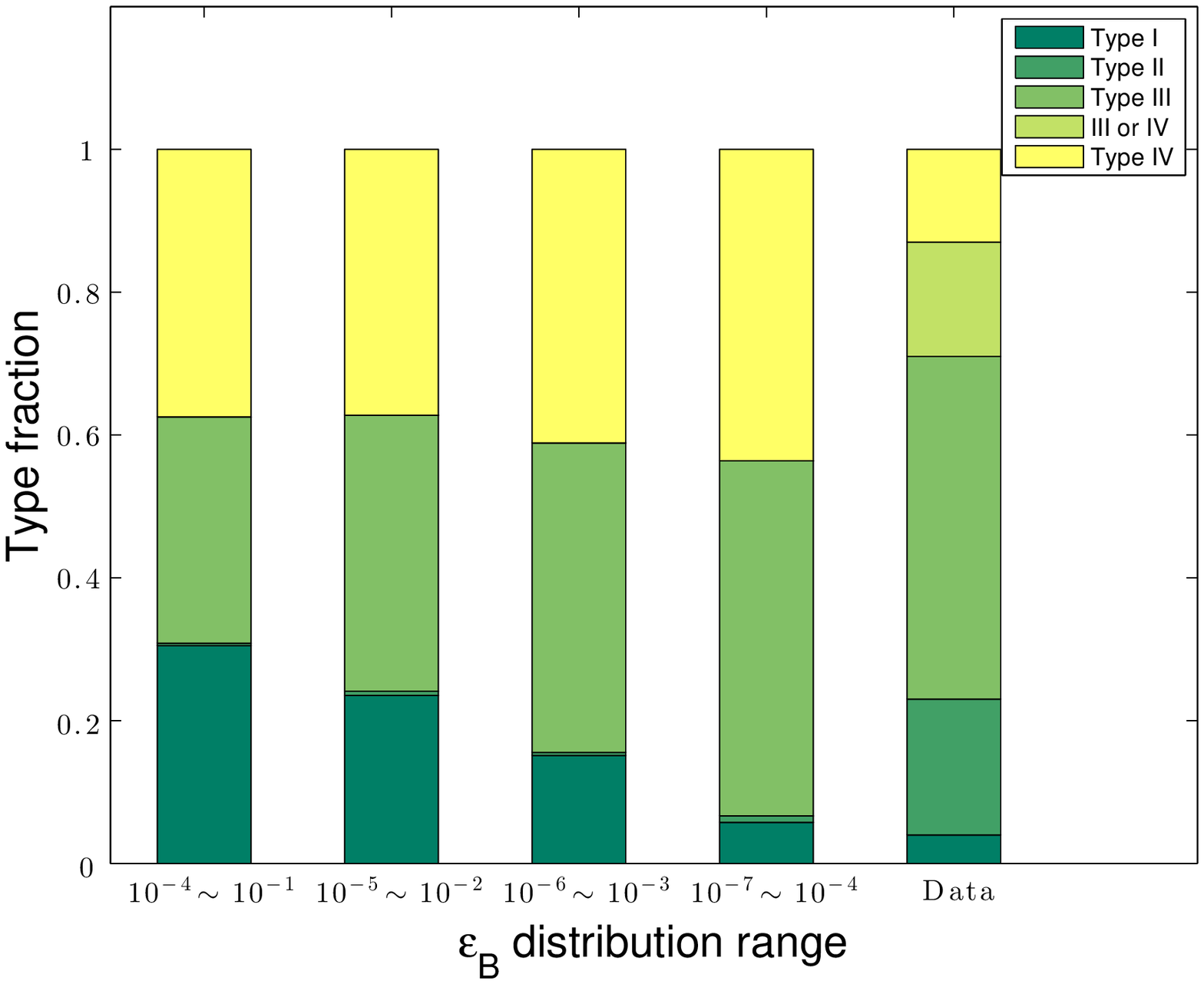}}
    \subfigure[$\epsilon_e^{f}=0.1,\RB=10,\alpha_E=-1$]{
    \label{fig:subfig:a}%% label for first subfigure
    \includegraphics[width=2.0in]{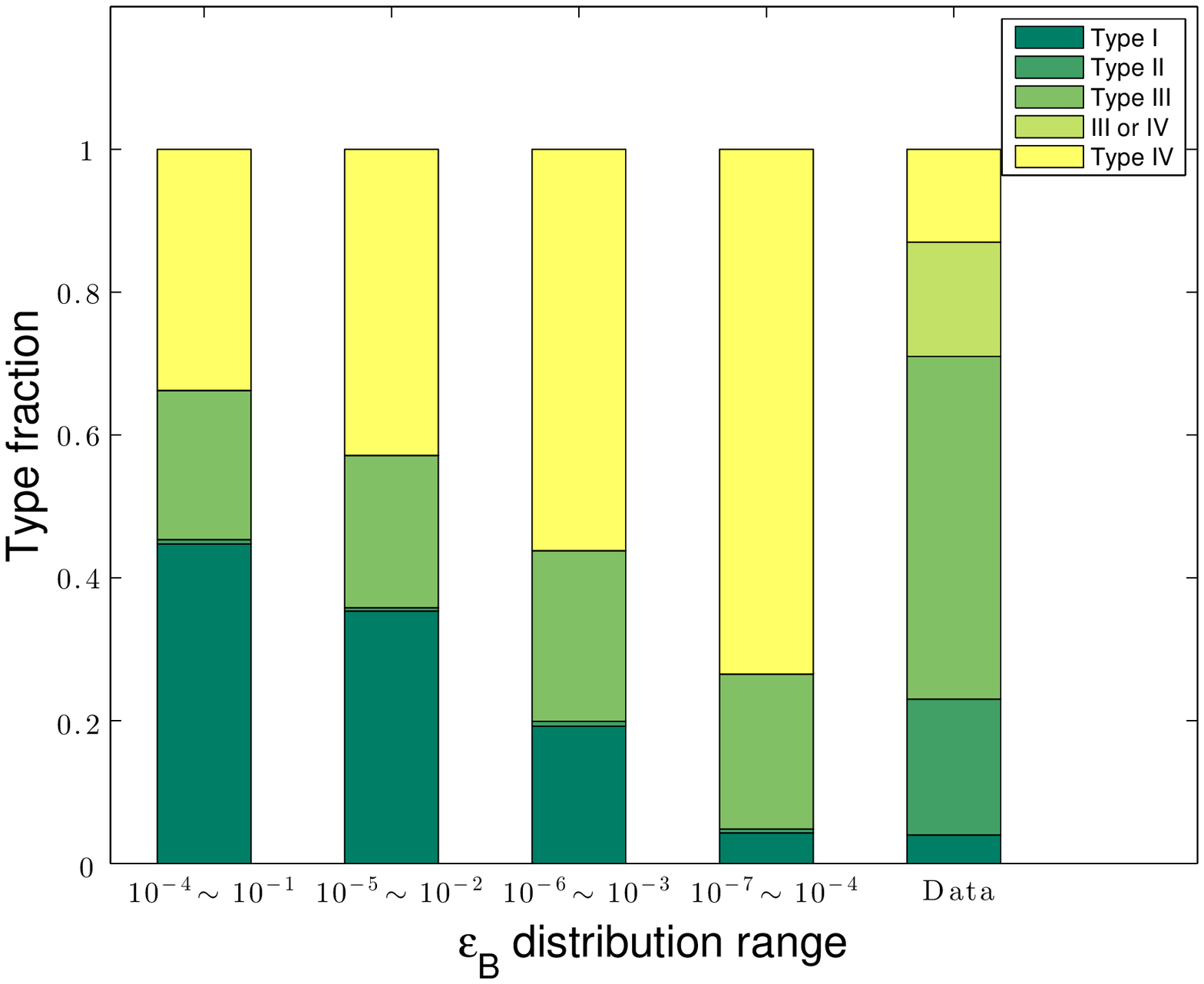}}   
    \subfigure[$\epsilon_e^{f}=0.1,\RB=10,\bar{\Delta}_0=10^{12}$]{
    \label{fig:subfig:a}%% label for first subfigure
    \includegraphics[width=2.0in]{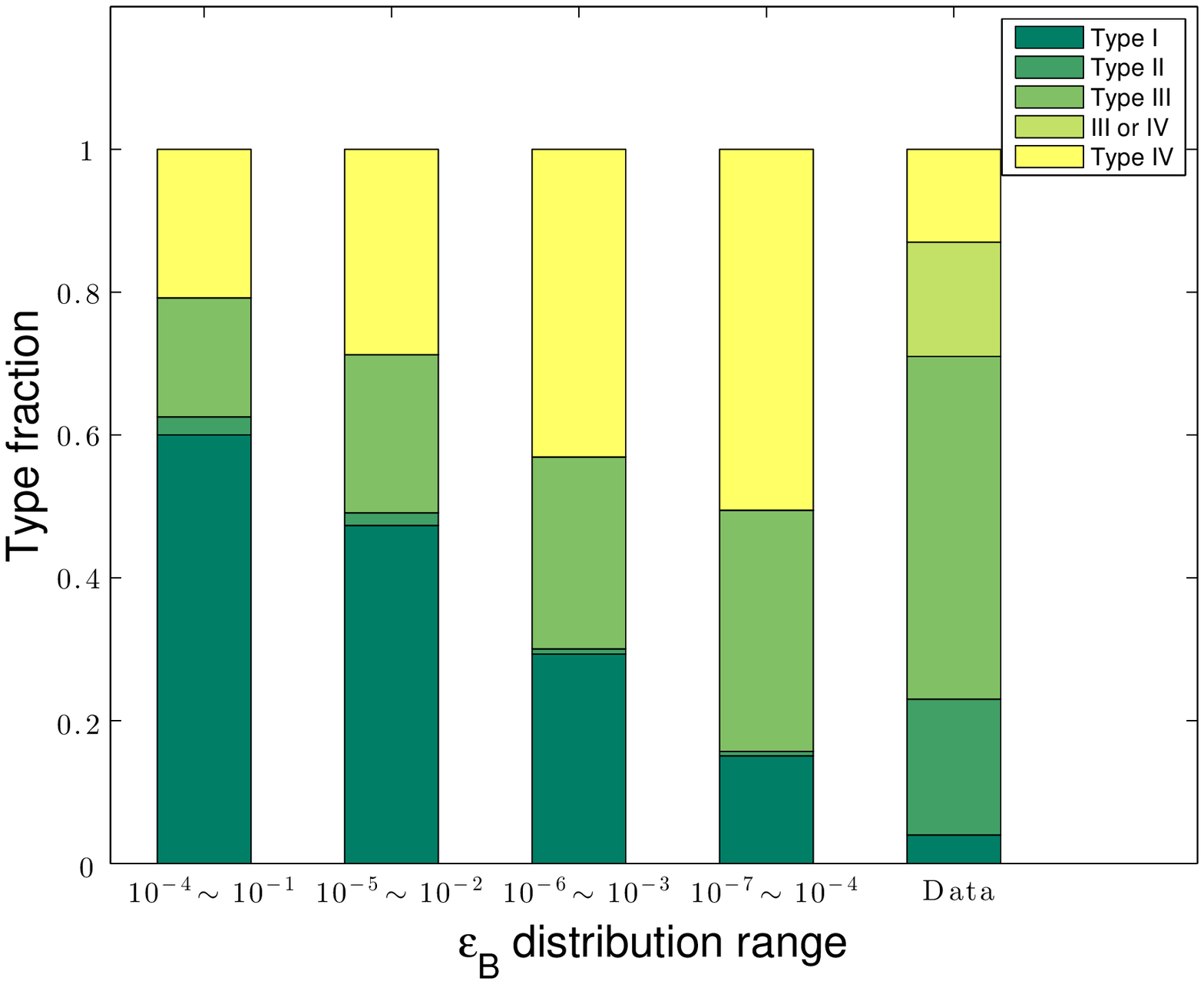}} \\   
\subfigure[$\bar{\epsilon}_{e}^{f}=0.01,\RBbar=1$]{
    \includegraphics[width=2.0in]{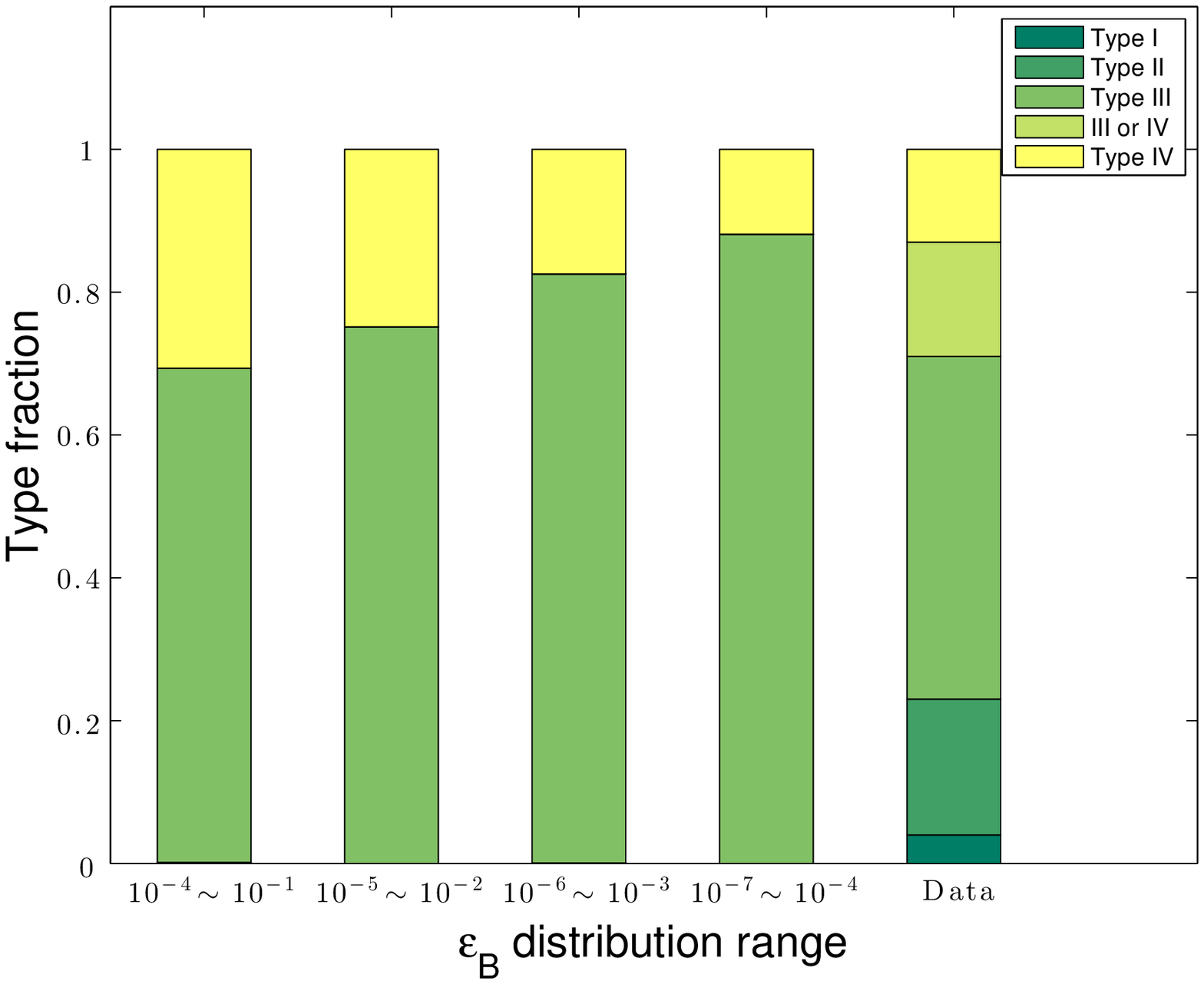}}
    \subfigure[$\bar{\epsilon}_{e}^{f}=0.01,\RBbar=10$]{
    \label{fig:subfig:a} %% label for first subfigure
    \includegraphics[width=2.0in]{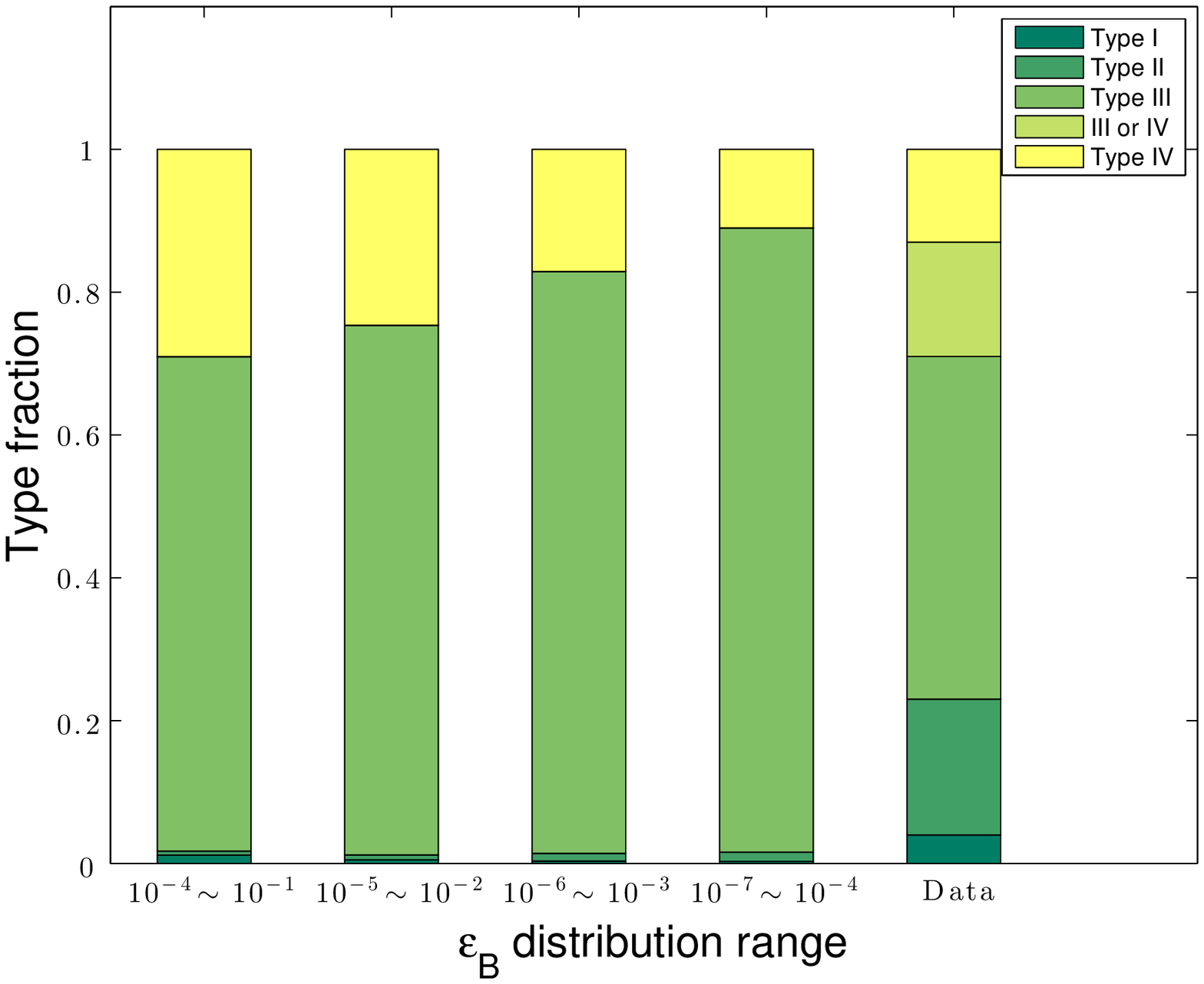}}
    \subfigure[$\bar{\epsilon}_{e}^{f}=0.01,\RBbar=100$]{
    \label{fig:subfig:a}%% label for first subfigure
    \includegraphics[width=2.0in]{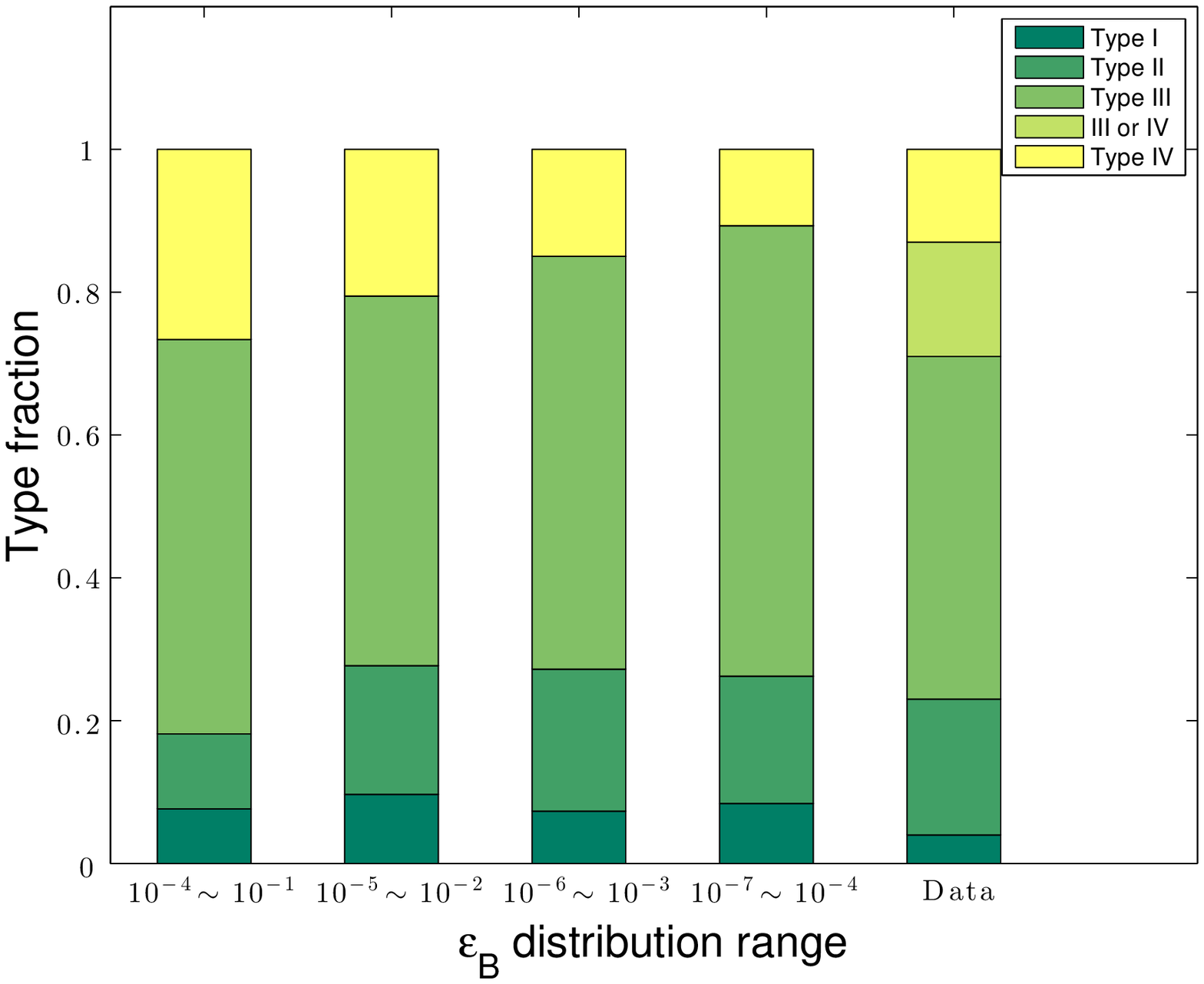}}\\
    \subfigure[$\bar{\epsilon}_{e}^{f}=0.001,\RBbar=1$]{
    \includegraphics[width=2.0in]{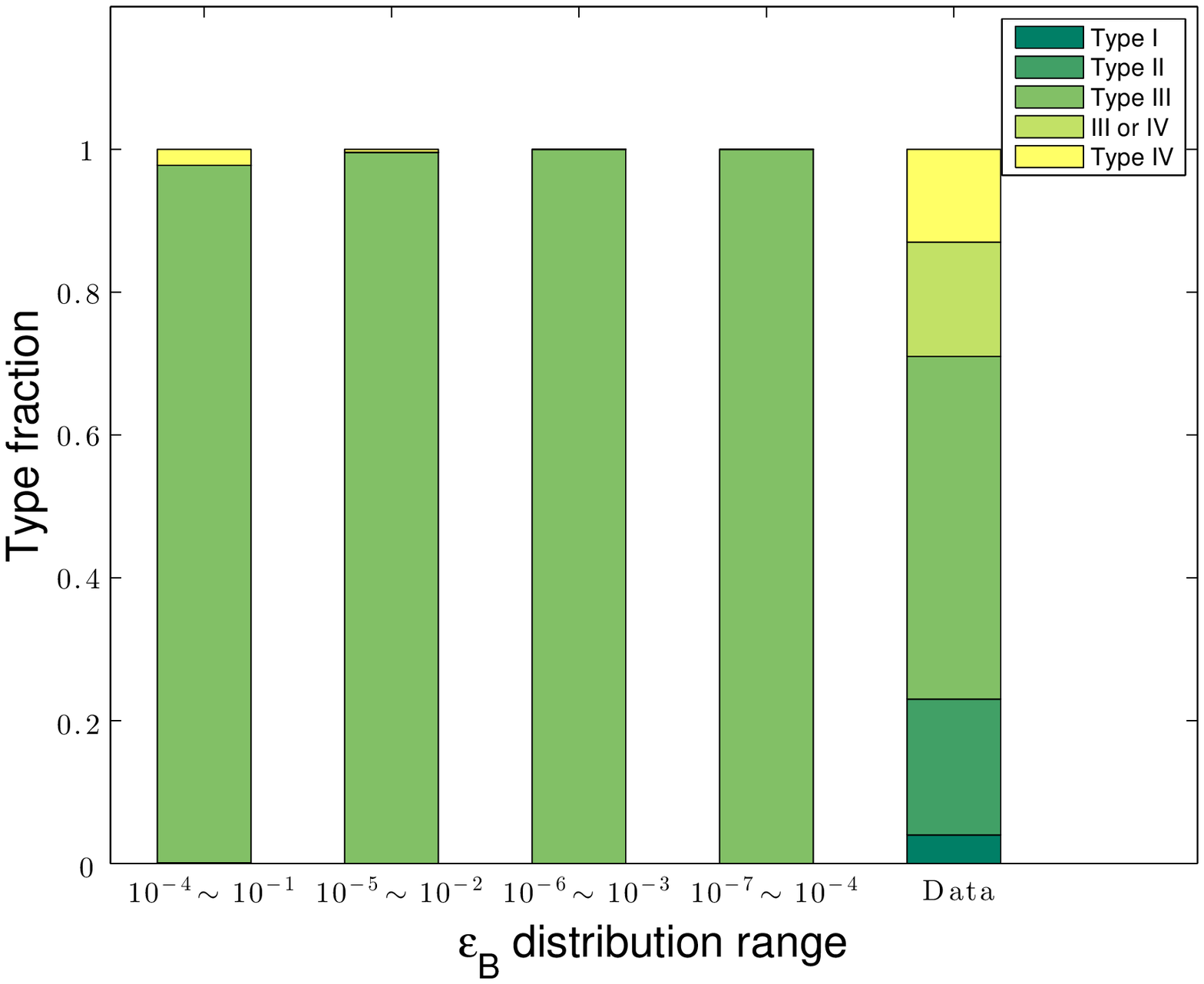}}
    \subfigure[$\bar{\epsilon}_{e}^{f}=0.001,\RBbar=10$]{
    \label{fig:subfig:a} %% label for first subfigure
    \includegraphics[width=2.0in]{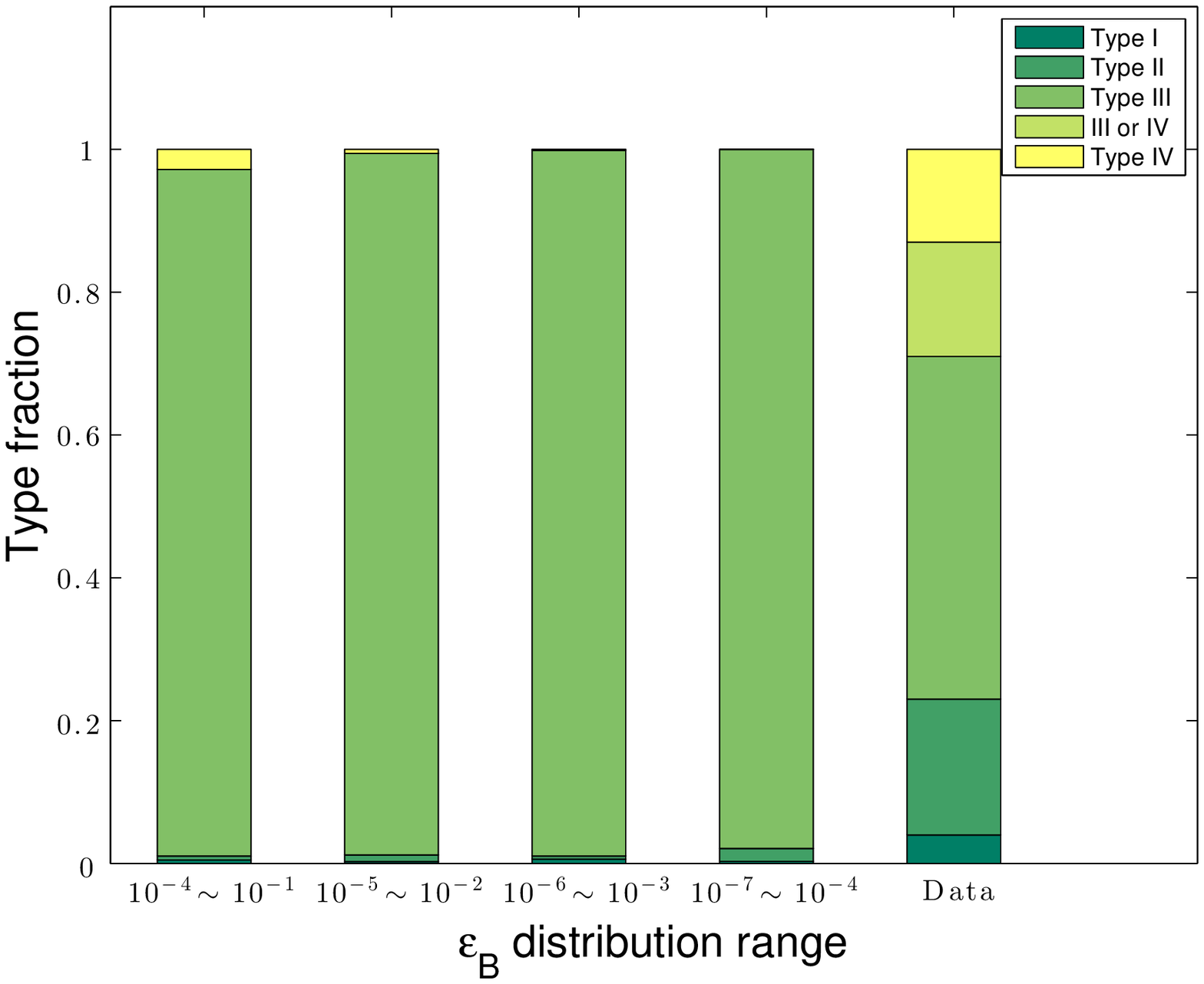}}
    \subfigure[$\bar{\epsilon}_{e}^{f}=0.001,\RBbar=100$]{
    \label{fig:subfig:a}%% label for first subfigure
    \includegraphics[width=2.0in]{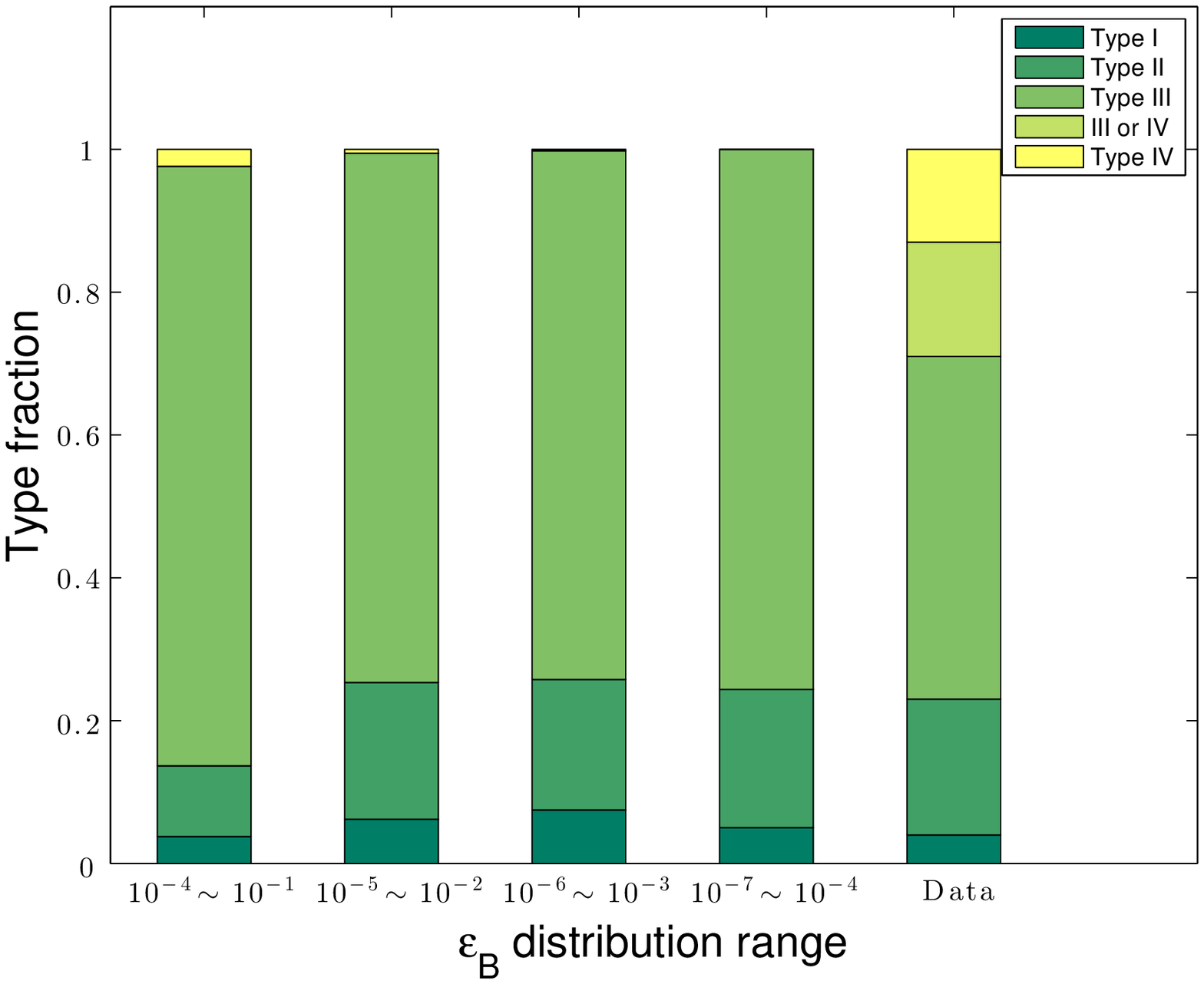}}\\
   \caption{Same a Figure \ref{fig:simu1} but for different parameters.}
           \label{fig:simu2}
            \end{figure*}    

\begin{itemize}
\item The fraction ratios between different lightcurve types depend sensitively on two key parameters, 
$\epsilon_e^{r,f}$ and $\RB$. The value of $\RB$ characterizes the balance between reverse shock 
dominated cases (I and II) and forward shock dominated cases (III and IV). 
Increasing $\RB$ can significantly increase the proportion of Type I and II. The value of 
$\epsilon_e^{r,f}$ essentially determines the internal coordination between Type I and II, 
or Type III and IV. Smaller $\epsilon_e^{r,f}$ gives more Type III and Type II. 
\item When $\bar{\epsilon}_{e}^{r,f}=0.1$, as shown in Figure \ref{fig:simu1}a-\ref{fig:simu1}c, $\RBbar$ 
should be larger than 10 but smaller than 100, otherwise the proportion of Type I and II is either 
too small or too large to reproduce the observational data. On the other hand, the observed fraction 
of Type II is much larger than Type I, which is in contrast with the simulation results. Varying 
the value of $\RBbar$ does not help to adjust the fraction ratio between Type I and II. 
\item Keeping $\epsilon_e^{r,f}$ of order of 0.1 and  $\RB$ of order of 10, we also checked if the 
observational results could be reproduced by varying other parameters. Since the simulation results 
depend sensitively on the value of $\epsilon_e^{r,f}$ and $\RB$, for better testing the effects of 
other parameters, we fix the value of $\epsilon_e^{r,f}$ as 0.1, the value of $\RB$ as 10, when the 
distribution functions of other parameters are being varied. The mean value of number density $\bar{n}$ is varied 
from 1 to 10 and 100; the mean value of electron index $\bar{p}$ is varied from 2.3 to 2.5 and 2.7; 
the distribution range of initial Lorentz factor $\G_0$ is varied from $50\sim300$ to $100\sim500$ and 
$50\sim500$; the power law index of kinetic energy distribution function is varied from 0.5 to 0.2 and 1; 
and the mean value of the initial shell width $\bar{\Delta}_0$ is varied from $10^{11}$ cm to $10^{12}$ cm. 
As shown in Figure \ref{fig:simu1}d-\ref{fig:simu2}c, varying the distributions of these parameters 
does not affect the results too much and hence does not help to solve the inconsistency of the ratio 
between Type I and Type II.
\item Fixing the distributions for all other parameter, the observations can be easily reproduced as long 
as the value of $\bar{\epsilon}_{e}^{r,f}$ is reduced by one order of magnitude i.e., $\bar{\epsilon}_{e}^{r,f}=0.01$ 
(when $\RBbar=100$ as shown in Figure \ref{fig:simu2}f). The constraint on $\epsilon_B^{f}$ is not strong, 
but smaller values of $\epsilon_B^{f}$ ranging from $10^{-6}$ to $10^{-2}$ seems to be more favorable.  
\item When $\bar{\epsilon}_{e}^{r,f}=0.001$, although the fraction of Type I and II could be consistent with the 
observations as long as $\RBbar$ is large enough, the fraction of Type IV is too small (or even completely 
disappear), which is inconsistent with the observations.
\end{itemize} 

In summary, the simulation results indicate that our morphological analysis for early optical afterglow is
able to efficiently constrain the microscopic parameters, e.g., $\epsilon_e^{r,f}$, $\epsilon_B^{f}$ 
and $\RB$. To reproduce the current observations, $\bar{\epsilon}_{e}^{r,f}=0.01$, $\RBbar=100$ and relatively 
smaller values of $ \epsilon_B^{f}$ is favored, which can be understood as follows: in the observational 
data, the fraction of Type II is larger than Type I, inferring that the peak of the forward shock emission 
is easily suppressed by the reverse shock component. On the other hand, the fraction of Type III is larger 
than Type IV, even when all the bursts of overlap type belong to Type IV. As illustrated in Figure 
\ref{fig:illustration}, both these items of observational evidence can be explained if the forward shock 
component is in the FS II case ($\nu_{m}^{f}(\tx)<\nuo$), which favors a smaller value of $\epsilon_e^{f}$. 
If $\epsilon_e^{f}$ becomes smaller, the forward shock emission in the optical band becomes stronger, so 
that a larger value of $\RB$ and a relatively smaller value of $\epsilon_B^{f}$ is required to maintain 
the balance between reverse shock dominated cases and forward shock dominated cases.

\section{Discussion}

A practical scheme of morphological analysis for GRB early optical afterglows and its ability to constrain 
afterglow parameters has been illustrated in the last two sections. We have applied this method to the 
currently available observational results, and have derived constraints on the relevant microscopic 
parameters. In the following, we will discuss some of the challenges facing this method and the caveats 
on our constraint results. 

The greatest challenge for the morphological analysis method arises from the sample selection. It is difficult 
to achieve the completeness of a certain sample, unless a sufficiently large number of triggered GRBs can
be rapidly followed-up in the optical band. On the other hand, systematic uncertainties could become large, 
and would be difficult to remove, if the afterglow follow-ups are obtained through different telescopes. 
A future dedicated facility with rapid response ability and wide field of view could help with these issues, 
and this is a key element in the Chinese-French mission SVOM, the Ground Wide Angle Cameras (GWACs) 
\citep{paul12}.

Another challenge comes from the process of assigning the observed lightcurves into relevant categories. 
To better identify the Type III and Type IV, sufficient data points in the rising phase are required, while 
to precisely distinguish Type I from Type II, observations in the decaying phase need to be dense enough. 
Multi-color observations during the follow-up phase are essential to address this challenge. 

The theoretical scheme for determining lightcurve categories is based on the standard synchrotron external 
shock model. Despite its great success, the standard model has some limitations that sometimes hinder a 
precise description of GRB afterglows. For instance, the real evolution of $\nu_m^{r,f}$ may deviates from 
a power-law behavior when $t$ is around $\tx$, so that both the reverse shock and forward shock lightcurves 
should have a smooth transition around the peak, especially when equal arrival time effects are considered. 
These deviations may affect our results over some limited range of parameter spaces, e.g., when $\Ff(\tfp)$ 
is close to $\Fr(\tfp)$. Such effects may average themselves out, as long as the simulated sample is large 
enough. In principle, one can use numerical simulations to calculate more precise lightcurves for given set 
of parameters, but this will dramatically increase the computation time while most of the calculations are 
redundant for the purpose of morphological analysis.

Due to the limitations of the current facilities, the sample selected in this work is still incomplete in 
some sense. As mentioned in section 2.3, only 114 swift bursts have optical follow-up within 500 s, and 
some of them are hard to classify because of lack of sufficient data points in the rising phase. The 
incompleteness may cause some uncertainty in the parameter constraint results, but the general tendency of 
our results should be reliable in order of magnitude, e.g., $\epsilon_e^{r,f}$ should be in order of 0.01 
and $\RB$ should be in order of 100. 

The analysis in this work is designed for a homogeneous interstellar medium. For GRBs occurring in a wind 
type environment \cite[e.g.][]{chevalier99}, the lightcurves are easy to distinguish from what is discussed 
here, and these may thus be excluded during the sample selection phase \citep{zhang03}. 

\section{Conclusion}

With decades of data accumulation and the prospects for future facilities, the GRB afterglow field is 
entering the era of big data. It is essential to find efficient methods to provide insights into the
general features of GRB afterglows from the study of large samples. In this work, we have developed
and implemented a morphological analysis method using Monte Carlo simulations, and find that such a 
morphological analysis applied to early optical afterglows can efficiently constrain the microscopic 
parameters, e.g., $\epsilon_e^{r,f}$, $\epsilon_B^{f}$ and $\RB$. To reproduce the current observational
data, $\epsilon_e^{r,f}$ distributed around $0.01$, $\RB$ distributed around $100$ and relatively smaller values of $\epsilon_B^{f}$ ranging from $10^{-6}$ to $10^{-2}$ are favored. If our interpretation is correct, two important implications can be 
inferred: 1) the preferred $\epsilon_e^{r,f}$ value is smaller than the commonly assumed value of 
$\epsilon_e^{r,f}=0.1$.  As a result, the same level of afterglow flux corresponds to a larger 
kinetic energy, which makes the measured radiative efficiency ($\eta=E_{\g}/(E_{\g}+E_{K})$, Lloyd-
Ronning \& Zhang 2004) lower than previously derived values. The internal shock models have suffered the 
criticism of a relatively low energy dissipation efficiency 
\citep{panaitescu99,kumar99,granot06b,zhang07a}, which is typically a few percent. A lower 
$\epsilon_e$ in the external shock would mitigate the ``low efficiency problem" of the 
internal shock model, if $\epsilon_e$ during the 
prompt emission phase (in the internal shocks) is large (say, $\sim 0.1$). This may be achievable if the 
relatively low $\epsilon_e$ as found in this paper is only relevant for extremely relativistic shocks, so that 
the mildly relativistic internal shocks may retain a relatively large $\epsilon_e$; 
2) values of $\RB=\epsilon_B^{r}/\epsilon_B^{f}\sim100$ correspond to $B_r/B_f \sim 10$, which is in 
agreement with the results of previous works which indicate a moderately magnetized baryonic GRB jet 
\citep{fan02,kumarpanaitescu03,zhang03,harrisonkobayashi13,japelj14}.

\acknowledgments{We thank an anonymous referee for a constructive report. This work is partially supported by NASA through grants NNX13AH50G and NNX14AF85G.}

%\bibliography{/Users/liris/D/Work/ms}

\end{document}